\documentclass[a4paper,11pt]{article}
\pdfoutput=1 
\usepackage{jheppub} 
\usepackage{amsmath,amssymb,epsfig,relsize,slashed,comment}
\allowdisplaybreaks

\definecolor{nicered}{rgb}{0.7,0.1,0.1}
\definecolor{nicegreen}{rgb}{0.1,0.5,0.1}
\definecolor{niceblue}{rgb}{0.0,0.1,0.7}
\hypersetup{colorlinks,citecolor=nicegreen,linkcolor=nicered,urlcolor=niceblue}

\def \beq{\begin{equation}}
\def \eeq{\end{equation}}
\def \bea{\begin{eqnarray}}
\def \eea{\end{eqnarray}}

\def \Z {\mathbb{Z}}

\usepackage{amsmath}
\usepackage{amsfonts}
\usepackage{amssymb}
\usepackage{graphicx}
\usepackage{mathtools}
\usepackage{braket}
\usepackage{enumerate}
\usepackage{bbold}
\usepackage{slashed}
\usepackage{feynmp-auto}
\usepackage{subfigure}
\usepackage{multirow}
\usepackage{booktabs}
\usepackage{soul}

\title{Split SIMPs with Decays}

\author[1,2]{Andrey Katz,}
\author[1]{Ennio Salvioni,}
\author[1]{and Bibhushan Shakya$\,$}

\affiliation[1]{Theoretical Physics Department, CERN, Esplanade des Particules 1, 1211 Gen\`eve 23, Switzerland}
\affiliation[2]{D\'epartement de Physique Th\'eorique and Center for Astroparticle Physics (CAP), Universit\'e de Gen\`eve, Quai Ernest-Ansermet 24, 1211 Gen\`eve 4, Switzerland}

\emailAdd{andrey.katz@cern.ch}
\emailAdd{ennio.salvioni@cern.ch}
\emailAdd{bibhushan.shakya@cern.ch}

\abstract{ We discuss a minimal realization of the strongly interacting massive particle (SIMP) framework. The model includes a dark copy of QCD with three colors and three light flavors. A massive dark photon, kinetically mixed with the Standard Model 
hypercharge, maintains kinetic equilibrium between the dark and visible sectors. One of the dark mesons is necessarily unstable but long-lived, with potential impact on CMB observables. We show that an approximate ``isospin'' symmetry acting on the down-type quarks is an essential ingredient of the model. This symmetry stabilizes the dark matter and allows to split sufficiently the masses of the other states to suppress strongly their relic abundances. We discuss for the first time the SIMP cosmology with sizable mass splittings between all meson multiplets. We demonstrate that the SIMP mechanism remains efficient in setting the dark matter relic density, while CMB constraints on unstable relics can be robustly avoided. We also consider the phenomenological consequences of isospin breaking, including dark matter decay. Cosmological, astrophysical, and terrestrial probes are combined into a global picture of the parameter space. In addition, we outline an ultraviolet completion in the context of neutral naturalness, where confinement at the GeV scale is generic. We emphasize the general applicability of several novel features of the SIMP mechanism that we discuss here.}

\preprint{CERN-TH-2020-107}

\begin{document} 

\maketitle

\section{Introduction}

In recent years, the strongly interacting massive particle (SIMP) framework~\cite{Hochberg:2014dra} has emerged as an attractive possibility for thermal dark matter, alternative to the traditional weakly interacting massive particle (WIMP) paradigm. The SIMP relic density is set by the freezeout of \mbox{$3\to 2$} self-annihilations, whose parametrics naturally point toward masses comparable to the strong scale, roughly between $10$~MeV and $1$~GeV, and strong coupling. If kinetic equilibrium between the dark matter and the Standard Model (SM) bath is maintained until the $3\to 2$ processes freeze out at $T \approx m_{\rm DM}/20$, the dark matter remains sufficiently cold to avoid conflict with structure formation bounds~\cite{Hochberg:2014dra}, which otherwise exclude~\cite{deLaix:1995vi} a completely secluded $3\to 2$ freezeout~\cite{Carlson:1992fn}.

The SIMP mechanism finds its most attractive realizations in the context of confining gauge theories with chiral symmetry breaking~\cite{Hochberg:2014kqa}, where the pseudo Nambu-Goldstone bosons (pNGBs) play the role of dark matter, and $3\to 2$ annihilations are mediated by the Wess-Zumino-Witten (WZW) action~\cite{Wess:1971yu,Witten:1983tw}. The pNGBs are also naturally characterized by strong $2\to 2$ self-scattering~\cite{Hochberg:2014dra,Hochberg:2014kqa}, which may help to address possible shortcomings of collision-less cold dark matter (CDM) on small scales (as reviewed for example in Ref.~\cite{Bullock:2017xww}). While such issues may eventually be resolved without modifying the CDM paradigm, dark matter that self-interacts with cross sections in the range $\sigma/m_{\rm DM} \sim 0.1\,$--$\,10$ cm$^{2}/\mathrm{g}$ is a very interesting possibility in this respect; see Ref.~\cite{Tulin:2017ara} for an extensive review.  

Existing studies of pNGBs as SIMP dark matter generally consider scenarios where all the pNGB mesons are (approximately) mass-degenerate, and the (dominant) component of dark matter is stable. The aim of this work is to study a minimal realization of the SIMP mechanism where variations of such properties can be explored. We consider a dark copy of the SM QCD, consisting of an $SU(N_c)$ gauge theory with $N_c = 3$ and \mbox{$N_f = 3$} light hidden quark flavors, which is the smallest $N_f$ that admits a WZW action~\cite{Witten:1983tw}, necessary for the realization of the $3\to2$ processes central to the SIMP mechanism. Hidden electromagnetism is gauged by a massive dark photon kinetically mixed with the SM hypercharge, providing a viable mediation mechanism that maintains kinetic equilibrium between the hidden and SM sectors~\cite{Lee:2015gsa,Hochberg:2015vrg}.

It was pointed out in Ref.~\cite{Berlin:2018tvf} that when $N_f$ is odd, one of the pNGBs is necessarily unstable. We show that, for typical parameters, the unstable meson $\eta$ decays to SM particles with lifetime comparable to the timescale of recombination, potentially leading to strong constraints from cosmic microwave background (CMB) anisotropies~\cite{Slatyer:2016qyl,Poulin:2016anj}. We show that an approximate $SU(2)_U$ global symmetry acting on the down and strange quarks, which we refer to as ``isospin,'' is a crucial component in this setup. This symmetry plays two key roles. First, it stabilizes the lightest multiplet, a triplet of dark mesons with mass in the $100\,$--$\,300$ MeV range. Second, it allows for separation between the masses of the up quark and the degenerate down-type quarks, which in turn raises the $\eta$ mass relative to the dark matter mass, suppressing the $\eta$ abundance to a level allowed by CMB measurements. In this regime, the masses of all $SU(2)_U$ multiplets are separated by similar, sizable amounts, raising another important question: namely, whether the $3\to 2$ freezeout remains effective even in this scenario of larger mass splittings. Our analysis provides a positive answer, opening up new parameter space for the SIMP mechanism. These results can be easily generalized to other models with odd $N_f$.  

In this context, we address another question that has remained surprisingly understudied in the literature: the absolute stability of SIMP dark matter. The neutral pion, which is one of the components of the dark matter triplet, decays through its mixing with $\eta$ induced by small breaking of isospin. We analyze quantitatively the sensitivity of current dark matter indirect detection searches to the order parameter of isospin breaking, finding that it should not exceed $O(10^{-5})$, and present the projected reach at future experiments. We combine these astrophysical and cosmological probes with laboratory tests of the dark photon mediator, painting a global picture of the parameter space. We emphasize that many of our results have broader applicability, beyond the minimal model adopted here. 

A brief discussion of the abundance and decays of unstable SIMP mesons was presented in Ref.~\cite{Hochberg:2018vdo}, albeit in a setup with $N_f = 4$. One of the main novelties of our work is a quantitative analysis of the SIMP cosmological history, focusing on larger mass splittings than previously considered in the literature and highlighting the key role of CMB anisotropy bounds on the unstable mesons. We work in pure chiral perturbation theory for the pNGBs, neglecting resonances such as the vector mesons, whose role has been extensively discussed in Ref.~\cite{Berlin:2018tvf} (see also Ref.~\cite{Choi:2018iit}).

As is characteristic of the SIMP framework, we find large dark matter self-scattering cross sections~\cite{Hochberg:2014kqa}. 
Our minimal choice of $N_c = 3$ leads to 
$\sigma / m_{\rm DM} \sim \mathrm{few}$ cm$^{2}/\mathrm{g}$, in some tension with bounds from the Bullet cluster~\cite{Randall:2007ph} and halo shapes~\cite{Rocha:2012jg,Peter:2012jh}. However, given the evolving status of the small-scale CDM puzzles, we believe that it would be premature to discard the minimal and theoretically appealing setup analyzed here. Furthermore, our results can serve as a useful basis for building models that feature smaller self-interaction cross sections.

Theoretical motivation for the scenario discussed here comes from neutral naturalness theories, such as the Twin Higgs~\cite{Chacko:2005pe}, which address the little hierarchy problem by introducing top partner particles that are not charged under SM color. Neutral naturalness models typically single out $N_c = 3$ in the hidden sector, as this allows the top partner to cancel the top quark loop correction to the Higgs mass. In addition, as argued in Ref.~\cite{Craig:2015pha}, two-loop naturalness considerations suggest that hidden color should confine at a scale similar to that of the SM QCD. Motivated by these arguments, we outline an embedding of the low-energy theory into a neutral naturalness model, along the lines of the vector-like Twin Higgs~\cite{Craig:2016kue}. In contrast with the ``Twin SIMPs'' setup of Ref.~\cite{Hochberg:2018vdo}, we do not introduce a full mirror copy of the SM, but instead propose a minimal construction for SIMP dark matter where the light dark quarks and the top partner(s) are more directly linked.

The remainder of our paper is structured as follows. In Section~\ref{sec:EFT} we introduce the effective theory for the hidden mesons, discussing in detail its symmetries, mass spectrum, and leading interactions. We outline ultraviolet completions in the framework of neutral naturalness in Section~\ref{sec:UV_compl}, which can be skipped by readers who are only interested in dark matter phenomenology. The lifetimes of the unstable mesons are calculated in Section~\ref{sec:mesondecay}. In Section~\ref{sec:cosmo_history} we discuss in detail the cosmological history, while signatures and constraints are presented in Section~\ref{sec:constraints}. Finally, we conclude in Section~\ref{sec:outlook} with a brief summary and outlook. Appendix~\ref{sec:BEs} provides complete Boltzmann equations for our setup. 

%%%%%%%%%%%%%%%%%%%%%%%%%%%%%%%%%%%%%%%%%%%%%%%%%%%%%%%%%%%%%%%%%%%%%%%%%%%%%%%%%%
%%%%%%%%%%%%%%%%%%%%%%%%%%%%%%%%%%%%%%%%%%%%%%%%%%%%%%%%%%%%%%%%%%%%%%%%%%%%%%%%%%%
%%%%%%%%%%%%%%%%%%%%%%%%%%%%%%%%%%%%%%%%%%%%%%%%%%%%%%%%%%%%%%%%%%%%%%%%%%%%%%%%%%%%
\section{Effective theory for the hidden mesons} 
\label{sec:EFT}
In this section we take a phenomenological approach and discuss the effective field theory (EFT) of hidden QCD, remaining agnostic about specific ultraviolet (UV) completions. Possible UV completions in the framework of neutral naturalness are addressed in Section~\ref{sec:UV_compl}. 

We assume an $SU(N_c)$ hidden color gauge group with $N_f = 3$ light hidden quark flavors. All our numerical results assume $N_c = 3$, although we occasionally comment on the effect of changing the number of colors. We also introduce a massive dark photon, kinetically mixed with the SM hypercharge, which keeps the hidden and SM sectors in kinetic equilibrium\footnote{An axion-like particle~\cite{Kamada:2017tsq,Hochberg:2018rjs} or vector mesons~\cite{Berlin:2018tvf} have also been studied as mediators for pNGB SIMP dark matter.} via elastic scattering at least until the $3\to 2$ processes among the dark matter (DM) particles freeze out~\cite{Lee:2015gsa,Hochberg:2015vrg}, as required for SIMP DM~\cite{Hochberg:2014dra}.\footnote{If the elastic scattering decouples before the $3\to 2$ processes, elastically-decoupling relic DM, or ELDER, can be realized~\cite{Kuflik:2015isi}.} The Lagrangian for the dark photon is
\begin{equation}
\mathcal{L}_g = - \frac{1}{4} \hat{F}_{\mu\nu} \hat{F}^{\mu \nu} + \frac{1}{2} m^2_{\hat{A}} \hat{A}_\mu \hat{A}^{\mu} + \frac{\varepsilon}{2} \hat{F}^{\mu\nu} B_{\mu\nu}\,.
\end{equation}
After diagonalization of the gauge kinetic and the mass terms, the physical dark photon $A'$ couples to the SM electromagnetic (EM) current at $O(\varepsilon)$. For fermions, this interaction reads $\varepsilon e c_w Q_f A'_\mu \bar{f}\gamma^\mu f$. We do not specify the origin of the $A^\prime$ mass, which could arise from the St\"uckelberg mechanism or from the coupling to a 
dark Higgs field. 

If the three quarks $u, d, s$ are light compared to the confinement scale, the pattern of low-energy spontaneous symmetry breaking is 
$SU(3)_L \times SU(3)_R \times U(1)_B \to SU(3)_V \times U(1)_B$, as in the SM. The $N_f^2 - 1 = 8$ Goldstone bosons are parametrized by
\begin{equation} 
\Sigma = \exp \Big( i \frac{\Pi}{f_\pi} \Big), \quad \Pi = 
\pi^a  \lambda^a , \quad \frac{\Pi}{\sqrt{2}} = 
\begin{pmatrix} \tfrac{1}{\sqrt{2}}\,\pi_3 + \tfrac{1}{\sqrt{6}}\,\pi_8 & \pi_+ & K_+  \\ 
\pi_- & - \tfrac{1}{\sqrt{2}}\, \pi_3 + \tfrac{1}{\sqrt{6}}\, \pi_8  & K_0 \\ 
K_- & \overline{K}_{0} & - \sqrt{\tfrac{2}{3}}\,\pi_8 \end{pmatrix} ,
\end{equation}
where $\lambda^a$ are the Gell-Mann matrices, satisfying $\mathrm{Tr}(\lambda^a \lambda^b) = 2\delta^{ab}$. Our normalization is such that 
$f_\pi^{\rm SM} \approx 92.4$~MeV, and the cutoff of the EFT is 
\beq \label{eq:def_Lambda}
\Lambda = 4\pi f_\pi\,,
\eeq
which is also our definition of the dark strong coupling scale. The terms of the chiral Lagrangian most relevant to our discussion are (see e.g. Ref.~\cite{Scherer:2002tk})
\begin{align}
\mathcal{L} \,=&\, \frac{f_\pi^2}{4}\, \mathrm{Tr} \big[ (D_\mu \Sigma)^\dagger 
D^\mu \Sigma \big] + \frac{Bf_\pi^2}{2} 
\mathrm{Tr}(M^\dagger \Sigma + \Sigma^\dagger M) + c f_{\pi}^4\, 
\hat{e}^2\, \mathrm{Tr}(\Sigma^\dagger \widehat{Q} 
\Sigma \widehat{Q}\hspace{0.2mm}) \nonumber \\
\,-&\,\frac{i \hat{e}^2 N_c}{48\pi^2} \epsilon^{\mu\nu\alpha\beta} 
\hat{F}_{\mu\nu} \hat{A}_\alpha \mathrm{Tr} \Big( \widehat{Q}^{\,2} \partial_\beta 
\Sigma \,\Sigma^\dagger - \widehat{Q}^{\,2} \partial_\beta \Sigma^\dagger \Sigma -
\frac{1}{2} \widehat{Q} \Sigma \widehat{Q} \partial_\beta \Sigma^\dagger + 
\frac{1}{2} \widehat{Q} \Sigma^\dagger \widehat{Q} \partial_\beta \Sigma \Big) 
\nonumber \\
\,+&\,  \frac{\hat{e} N_c}{48\pi^2} \epsilon^{\mu\nu\rho\sigma}  
\hat{A}_\mu \mathrm{Tr} \Big( \widehat{Q} \partial_\nu \Sigma\, 
\Sigma^\dagger \partial_\rho \Sigma\, \Sigma^\dagger \partial_\sigma \Sigma\, 
\Sigma^\dagger + \widehat{Q} \Sigma^\dagger \partial_\nu \Sigma\, 
\Sigma^\dagger \partial_\rho \Sigma\, \Sigma^\dagger \partial_\sigma \Sigma \Big)   
\nonumber \\
\,+&\, \frac{N_c}{240\pi^2 f_\pi^5}\, \epsilon^{\mu\nu\rho\sigma} \mathrm{Tr}\, 
\big( \Pi \partial_\mu \Pi \partial_\nu \Pi \partial_\rho 
\Pi \partial_\sigma \Pi \big)\,,  \label{eq:ChPT}
\end{align}
where $\Sigma \to L \Sigma R^\dagger$, and likewise for the quark mass spurion $M$. 
The interactions with external vector fields are described by spurions with formal 
transformation properties $\widehat{Q}_L \to L \widehat{Q}_L L^\dagger$ and $\widehat{Q}_R \to R \widehat{Q}_R R^\dagger$, but in Eq.~\eqref{eq:ChPT} we have 
already set $\widehat{Q}_L = \widehat{Q}_R = \widehat{Q}$ as appropriate to 
describe the coupling to hidden EM. The covariant derivative of $\Sigma$ reads $D_\mu \Sigma = \partial_\mu \Sigma - i \hat{e} \hat{A}_\mu [\widehat{Q}, \Sigma]$. The second and third lines of $\mathcal{L}$ display pieces of the WZW action that arise due to the presence of the gauge fields. In particular, the second line is relevant to the calculation of $\pi \to A^{\prime *}A^{\prime *}$ decays, whereas the third line is responsible for the semi-annihilations $\pi \pi \to \pi A^{\prime}$. The fourth line shows the piece of the WZW action that controls $3\to 2$ scattering among the mesons; it is nonzero only if all five participating mesons are different. 

The SM values of the electric charges are
\beq 
Q = \frac{1}{2} \Big(\lambda^3 + \frac{\lambda^8} {\sqrt{3}} \Big) = \frac{1}{3}\,
\mathrm{diag}\,(2, - 1, -1)\,.
\eeq 
More generally, assuming the underlying microscopic theory to be vector-like 
(so that the dark baryon number $B$ is anomaly-free), we have the freedom to gauge a linear combination of $Q$ and $B$. In particular, gauging
\beq 
Q' = Q - \frac{B}{2} = \frac{1}{2}\,  \mathrm{diag}\,(1, - 1, -1)\,
\eeq 
results in the vanishing of the axial-vector-vector (AVV) anomalies as a 
consequence of \mbox{$\mathrm{Tr}(Q^{\prime\,2} \lambda^a) =0$}, ensuring that 
even singlet mesons do not decay through anomalous diagrams. In this 
paper we consider both possibilities, $\widehat{Q} = Q$ and $\widehat{Q} = Q'$, 
for the charges.

We will require that the semi-annihilations $\pi\pi \to \pi A'$ be sub-leading to $3\to 2$ processes during DM freezeout (see Section~\ref{sec:semiann_ann}), which imposes a lower bound $m_{A'}\gtrsim 2\hspace{0.1mm} m_\pi$ on the mass of the dark photon~\cite{Hochberg:2015vrg}.

\subsection{Mass spectrum}
\label{sub:spectrum}
Setting the mass spurion to its physical value $M=\mathrm{diag}\,(m_u, m_d, m_s)$ 
gives the following pNGB masses, 
\begin{equation}
m^2_{\pi_\pm} = B(m_u + m_d) + \Delta m^2_{\rm em}\,, \qquad m^2_{K_\pm} = B(m_u + m_s) + \Delta m^2_{\rm em}\,, \qquad m^2_{K_0, \overline{K}_0} = B(m_d + m_s), 
\end{equation}
where $\Delta m^2_{\rm em} =  2 c \, \hat{e}^2f_\pi^2$ is the electromagnetic 
correction, with $c$ being a constant. %We will see momentarily that in the dark sector $c$ is naturally much smaller than $1$. 
The $\pi_3$ and $\pi_8$ mix as
\begin{equation}
\quad M^2_{\pi_3 \pi_8} = B \begin{pmatrix} m_u + m_d & \frac{1}{\sqrt{3}} (m_u - m_d) \\  \;\;\frac{1}{\sqrt{3}} (m_u - m_d) & \;\;\frac{1}{3}(m_u + m_d + 4m_s) \end{pmatrix},
\end{equation}
where $B$ is an a-priori unknown parameter of order the strong scale $\Lambda$, defined in Eq.~\eqref{eq:def_Lambda}. 
In the SM, the $\pi_0$ mass gives $B_{\rm SM} = 2.7$~GeV ($B$ has dimension of mass, so the natural dimensionless parameter is 
$b = B/\Lambda$ with $b_{\rm SM} = 2.3$) and the $\pi_\pm$--$\,\pi_0$ mass difference gives $c_{\rm SM} = 0.8$. In the following 
we simply take $b\sim O(1)$, whereas the EM correction $c$ has a different scaling 
compared to the SM, since we assume the dark photon is heavy, $m_{A'} \gtrsim 2 m_\pi > \Lambda$. We estimate $c\sim (\Lambda^2/m_{A'}^2) \log\, (m_{A'}^2 / \Lambda^2)$ (see e.g.~Ref.~\cite{Balkin:2018tma}), 
implying that this correction is small throughout our parameter space.   

For phenomenological reasons that will become clear momentarily, we focus on the scenario where the up quark is moderately heavier than the down-type quarks, which are approximately degenerate. Assuming $m_u + m_d > 2m_s$, the above matrix is diagonalized as
\begin{equation} 
\begin{pmatrix} \pi_3 \\ \pi_8 \end{pmatrix} = R(\varphi) \begin{pmatrix} \eta \\ \pi_0 \end{pmatrix}, \;\; \tan\, 2\varphi \,= \,\frac{\sqrt{3}\,(m_u - m_d)}{2 m_s - m_u - m_d}\,, \;\; R(\varphi)^T M^2_{\pi_3 \pi_8} R(\varphi) = \mathrm{diag}\,(m^2_{\eta}, m^2_{\pi_0})\,,
\end{equation}
with
\begin{equation}
m^2_{\pi_0, \eta} = \frac{2B}{3} \Big(m_u + m_d + m_s \mp \sqrt{ m_u^2 + m_d^2 + m_s^2 - m_u m_d - m_u m_s - m_d m_s }\,  \Big).
\end{equation}
The rotation matrix is defined as 
$R(\varphi) \equiv \begin{pmatrix} c_\varphi & s_\varphi \\ - s_\varphi & c_\varphi
\end{pmatrix}$, employing short-hand notations for sine and cosine that we use throughout the paper (in particular, $s_w$ and $c_w$ refer to the weak mixing 
angle). When the mass splittings are neglected, we denote the common octet mass simply as $m_\pi$.

As we have anticipated, we focus on a scenario where the down-type quarks are nearly degenerate due to an approximate $SU(2)$ symmetry. Hence, we parametrize the dark quark masses as
\begin{equation} \label{eq:quark_masses}
(m_u, m_d, m_s ) = (m + \Delta m, m, m + dm),
\end{equation}
where $\Delta m > 0$ is a sizable splitting of $ O(0.1\,$--$\,1)m$, whereas 
$dm$ can have either sign but is very small, $|dm| / \Delta m \ll 1$. The meson masses in this limit are
\begin{align}
\qquad\qquad&\,m^2_{\pi_0} \simeq B \Big(2m + dm - \frac{(dm)^{2}}{4\Delta m} \Big) 
\quad \lesssim \quad m^2_{K_0, \overline{K}_0} = B (2m + dm) \nonumber  \\
\,<\; m^2_{\pi_\pm} =&\, B (2m + \Delta m) + \Delta m^2_{\rm em} \quad 
\lesssim\quad  m^2_{K_\pm} = B (2m + \Delta m + dm) + \Delta m^2_{\rm em} \nonumber \\
 &\qquad\quad\,<\;\; m^2_{\eta} \simeq B \Big(2m + \frac{4\Delta m}{3} + \frac{dm}{3} +
 \frac{(dm)^2}{4\Delta m}\Big)\,, \label{eq:meson_masses}
\end{align}
where in the ordering of the charged mesons we have assumed $dm > 0$ for concreteness. 
The mixing angle between the $\pi_0$ and $\eta$ reads
\begin{equation}
s_\varphi = - \frac{1}{2} \Big( 1 + \frac{3\delta}{4}  + O(\delta^2) \Big), 
\qquad c_\varphi = \frac{\sqrt{3}}{2} \Big( 1 - \frac{\delta}{4}  + O(\delta^2) \Big),
\end{equation}
where we have defined 
\begin{equation} \label{eq:delta_param}
\delta \equiv \frac{dm}{ \Delta m }~.
\end{equation}
This parameter measures the strength of the $SU(2)_U$ breaking compared to the chiral $SU(3)$ breaking; however, since in most of our discussion we consider $\Delta m / m \sim O(1)$, we can take $\delta$ as effectively measuring the strength of isospin breaking. We also define the relative splittings
\begin{equation}
\Delta_{\pi, \eta} \equiv \frac{ m_{\pi_+, \eta} - m_{\pi_0}}{m_{\pi_0}}\,,
\end{equation}
which will be used frequently in later discussions.

In the isospin-symmetric limit $\delta = 0$, the mesons transform as
$\mathbf{3}_0$, $\mathbf{2}_{\pm 1}$ and $\mathbf{1}_0$ under the 
$SU(2)_U \times U(1)_{Q}$ 
symmetry, which is exactly preserved by the quark masses and by
electromagnetism.\footnote{This spectrum was previously considered in a 
very different 
regime, with $O(100)$~GeV dark meson masses~\cite{Beauchesne:2019ato}.} In particular, the $SU(2)_U$ triplet, which constitutes the DM, is formed by $(\pi_0, K_0, \overline{K}_0)$, 
the doublets are $(\pi_+, K_+)$ and $(\pi_-, K_-)$, and the singlet is $\eta$.
In the SM context the $SU(2)_U$ is often called U-spin, justifying its name, but for simplicity we refer to it as ``isospin.'' We assume
that $U(1)_Q$ is a good low-energy global symmetry, despite a heavy $A'$, implying, in particular, that all the charged pions and kaons are (almost) 
mass-degenerate and stable. This is guaranteed if $m_{A'}$ arises from a St\"uckelberg mechanism, whereas it could be a good approximation in certain realizations of a dark Higgs mechanism.

In Table~\ref{tab:mesonproperties} we present the explicit field contents of the dark mesons in the isospin-symmetric limit, as well as an overview of other salient properties, some of which will be analyzed later. With the exception of the $\eta$, for $\delta = 0$ the mesons cannot decay at any order in the chiral Lagrangian. As we discuss in Section~\ref{sec:mesondecay}, 
the $\eta$ does decay even in the isospin-symmetric limit, while the $\pi_0$ decays via small isospin-breaking effects. A stable $SU(2)$ triplet of SIMP DM mesons composed of hidden $d,s$ quarks was previously considered in Ref.~\cite{Hochberg:2018vdo}, albeit in a theory also containing light and degenerate $u,c$ quarks, leading to an extended pattern of chiral symmetry breaking.
\renewcommand{\tabcolsep}{5pt}
\begin{table}[t]
\centering
\begin{tabular}{c|c|c|c|l}
%\begin{tabular}{p{0.8cm}|p{1.3cm}|p{1.7cm}|p{4.5cm}|p{4.7cm}}
\multirow{2}{*}{rep.} & \multirow{2}{*}{meson} & quark content & mass & \multirow{2}{*}{stable?}\\ 
&  & $\simeq$ & $\simeq$   & \\\hline
\multirow{3}{*}{$\mathbf{1}_0$} & \multirow{3}{*}{$\eta$}   &  \multirow{3}{*}{$\frac{2 u\bar{u} - d\bar{d} - s\bar{s}}{\sqrt{6}}$} & \multirow{3}{*}{$m_{\pi_0}\big(1 + \frac{4}{3} \Delta_\pi \big)$} &  no; decays via AVV anomaly \\ 
 &   &   &  &  or higher-order operators \\
  &   &   &  & (Section~\ref{subsec:etadecay})\\\hline 
\multirow{2}{*}{$\mathbf{2}_{\pm 1}$} & $K_+, K_- $                   &  $u \bar{s}, \bar{u} s$ & $m_{\pi_0}(1 + \Delta_\pi + \Delta_\pi \delta)$ & \multirow{2}{*}{yes; charged under $U(1)_Q$} \\ 
      &       $\pi_+ , \pi_- $        &  $u \bar{d}, \bar{u} d$ &  $m_{\pi_0}(1 + \Delta_\pi)$ &  \\ \hline
\multirow{5}{*}{$\mathbf{3}_0$}     &      \multirow{2}{*}{$K_0,\,\overline{K}_0$}     & \multirow{2}{*}{$d\bar{s},\,\bar{d}s$} & \multirow{2}{*}{$m_{\pi_0}\big( 1 + \frac{1}{4}\Delta_\pi \delta^2\big)$} &  yes; charged under residual $U(1)$  \\
     &       &   &   & (Section~\ref{subsec:dmdecays})  \\ 
      &     \multirow{3}{*}{$\pi_0$}      & \multirow{3}{*}{$\frac{ d\bar{d} - s\bar{s} }{\sqrt{2}}$} & \multirow{3}{*}{$m_{\pi_0}$} &\rule{0pt}{1.25em}stable in isospin-symmetric limit; \\
     &         &  & & decays via $\propto \delta$ mixing with $\eta$ \\
          &         &  & & (Section~\ref{subsec:dmdecays}) \\
\end{tabular}
\caption{Summary of the properties of the hidden mesons, ordered by decreasing mass (assuming $\delta > 0$). The first column lists the representation under the $SU(2)_U \times U(1)_{Q}$ global symmetry. The quark contents correspond to the isospin-symmetric limit. Some of the masses are approximate; complete expressions are given in Section~\ref{sub:spectrum}.}
\label{tab:mesonproperties}
\end{table}

Note that choosing $\Delta m < 0$ in Eq.~\eqref{eq:quark_masses} leads to a spectrum where the singlet is the lightest meson 
and therefore comes to dominate the hidden sector abundance after freezeout. As the singlet is unstable even in the isospin-symmetric limit, with a lifetime much longer than one second but not arbitrarily long due to the requirement of thermalization between the hidden and SM sectors, this possibility is not viable.

The SIMP mechanism typically requires rather large values of $m_\pi / f_\pi$, and in this work we consider $m_{\pi_0}/f_\pi \sim 8$\,--$\hspace{0.2mm}10$ (for reference, in the SM $m_K/f_\pi \approx 5.4$). As the strongly-coupled regime is approached, higher-order corrections in chiral perturbation theory become important~\cite{Hansen:2015yaa}. Furthermore, when the pNGBs are heavy, additional resonances can play an important role in the dynamics. In 
particular, Refs.~\cite{Berlin:2018tvf,Choi:2018iit} explicitly introduced vector mesons in the effective theory, showing that this opens up additional parameter space for hidden meson DM. In general, the effects of the vector mesons can be neglected as long as their masses satisfy $m_{V} > 2 m_\pi\,$; otherwise, $V\to \pi \pi$ decays are kinematically closed and the semi-annihilations $\pi\pi \to \pi V$ followed by $V\to \mathrm{SM}$ decays play an important role in the cosmological evolution~\cite{Berlin:2018tvf}. Here, we assume $m_{V} > 2 m_\pi$ holds and focus on a minimal framework, neglecting resonances.

%%%%%%%%%%%%%%%%%%%%%%%%%%%%%%%%%%%%%%%%%%%%%%%%%%%%%%%%%%%%%%%%%%%%%%%%%%%%%%%%%
%%%%%%%%%%%%%%%%%%%%%%%%%%%%%%%%%%%%%%%%%%%%%%%%%%%%%%%%%%%%%%%%%%%%%%%%%%%%%%%%%%
%%%%%%%%%%%%%%%%%%%%%%%%%%%%%%%%%%%%%%%%%%%%%%%%%%%%%%%%%%%%%%%%%%%%%%%%%%%%%%%%%%%
\section{Ultraviolet completions in neutral naturalness} 
\label{sec:UV_compl}
We now outline possible UV completions of the chiral Lagrangian presented in the previous section in the context of neutral naturalness. The reader who is only interested in the phenomenological aspects of our work, or perhaps favors a different class of completions, may choose to proceed directly to Section~\ref{sec:mesondecay} without loss of continuity. 

The minimum requirements for a neutral natural theory that realizes the meson spectrum discussed in Section~\ref{sec:EFT} are:
\begin{itemize}
    \item two light and degenerate down-type quarks;
    \item one moderately heavier (but still lighter than the confinement scale) 
    up-type quark;
    \item one top partner that cancels the quadratic UV sensitivity of the 
    Higgs mass induced by the top Yukawa coupling.
\end{itemize}
The simplest construction with these characteristics can be built along the lines 
of the vector-like Twin Higgs~\cite{Craig:2016kue},
\beq\label{eq:model0} 
 - \mathcal{L}_f = - \hat{y}_t \widehat{H}^T\hspace{-0.75mm}\epsilon Q u^c + 
 \hat{y}^\prime_t Q^c \epsilon \widehat{H}^\ast u + 
 \hat{y}_b \widehat{H}^\dagger Q d^c + 
 \hat{y}_b^\prime Q^c \widehat{H} d + \mathrm{h.c.}\,,
\eeq
where $\epsilon = i \sigma^2$. The vector-like masses for the fermion fields are assumed to be small 
perturbations to the Yukawas and not written explicitly. The fields have the following charges 
under the twin $SU(2)_L \times U(1)_Y$ symmetry,
\beq 
 \widehat{H} \sim \mathbf{2}_{1/2}\, , \quad Q = 
 \begin{pmatrix} t \\ b \end{pmatrix} \sim \mathbf{2}_{1/6}\, , 
 \quad Q^c = ( t^c\,\, b^c ) \sim \mathbf{\bar{2}}_{-1/6}\,, \quad u,u^c 
 \sim \mathbf{1}_{\pm 2/3}  \,, \quad d,d^c \sim \mathbf{1}_{\mp 1/3}\,,
\eeq
where we have assumed the same hypercharge assignments as in the SM. The fields $Q, u, d$ transform as triplets of the twin $SU(3)_c$, and $Q^c, u^c, d^c$ transform as anti-triplets. We assume that the global (approximate) $SU(4)$ is non-linearly realized and the radial mode is heavier than the cutoff. Therefore, in unitary gauge the twin Higgs doublet is parametrized as
\beq
\widehat{H} = \begin{pmatrix} 0 \\ \tfrac{1}{\sqrt{2}} f \cos (h /f) \end{pmatrix} \qquad {\rm with} \qquad 
f \sin  (\langle h \rangle / f) = v \simeq 246\;\mathrm{GeV}.
\eeq
If $\hat{y}_t = y_t$ is enforced by a 
$\Z_2$ symmetry and $\hat{y}_t^\prime, \hat{y}_b, \hat{y}_b^\prime \ll  y_t$, 
the quadratic correction to the Higgs mass from the SM top loop is canceled by the 
twin top with mass $\simeq (f/v) m_t$. In addition, gauging the twin 
$SU(2)_L$ with $\Z_2$-symmetric coupling $\hat{g} = g$ ensures that the leading 
gauge corrections to the Higgs mass cancel as well. This is a simple vector-like 
Twin Higgs scenario with light twin quarks, leading to chiral symmetry breaking in the hidden sector (see Ref.~\cite{Freytsis:2016dgf} for similar ideas). 

Naively, taking 
$\hat{y}_t^\prime \gtrsim \hat{y}_b = \hat{y}_b^\prime$ in Eq.~\eqref{eq:model0} seems to provide exactly the light quark spectrum we desire. However, the fact that $y_t = \hat{y}_t \gg \hat{y}_t^\prime$ leads at one loop 
level to different renormalization of the two down-sector Yukawas via diagrams involving the twin $W$. The natural value of the isospin-breaking parameter is therefore one loop factor, significantly exceeding $\delta \lesssim 10^{-5}$, which is necessary to render the $\pi_0$ sufficiently long-lived to be a viable DM component. Thus, in this model an acceptably small $\delta$ can be achieved only at the price of fine tuning. 

This problem can be fixed by requiring that $\hat{y}_t = \hat{y}_t^\prime$, thus considering instead the Lagrangian
\beq\label{eq:model1} 
 - \mathcal{L}^\prime_f = - \hat{y}_t \widehat{H}^T\hspace{-0.75mm}\epsilon Q u^c + 
 \hat{y}_t Q^c \epsilon \widehat{H}^\ast u + 
 \hat{y}_b \widehat{H}^\dagger Q d^c + 
 \hat{y}_b Q^c \widehat{H} d + M_{u'} u^{\prime c} u'  + \mathrm{h.c.}\,.
\eeq
In this case the little hierarchy problem can be solved if $\hat{y}_t = y_t / \sqrt{2}$, with two heavy, degenerate top partners of mass $\simeq (f/v)m_t / \sqrt{2}$ now canceling the SM top loop. We envisage that a UV completion of 
Eq.~\eqref{eq:model1} may be constructed with Orbifold Higgs methods~\cite{Craig:2014aea,Craig:2014roa}, although we do not attempt to do so here. We have introduced an additional vector-like 
fermion $u^\prime, u^{\prime c}$ in order to have a light up-type quark in the spectrum. Our phenomenological study shows that only a 
moderate coincidence of scales, $M_{u'} \gtrsim \hat{y}_b f$ within an $O(1)$ factor, 
is necessary for a viable SIMP scenario. Equation~\eqref{eq:model1} provides 
a technically natural setup that preserves the isospin $SU(2)_U$. Additional masses and interactions 
\beq
 - \delta \mathcal{L}^\prime_f =  M_Q Q^{c} Q + M_u u^c u + M_d d^c d +
 (\mathrm{Yukawas}\;\mathrm{involving}\;u^{\prime c} , u') + \mathrm{h.c.}
\eeq
can then provide small breaking of 
isospin.\footnote{A priori, another possibility to obtain the desired spectrum is to 
assume that the Yukawa couplings are negligible in the down sector and the leading contribution to the down-type quark masses comes from 
$M_Q = M_d = M$. In this case we have a single top partner with $\hat{y}_t = y_t$, the required 
coincidence of scales for the up quark mass reads $\hat{y}_t^\prime f \gtrsim M$, and $u^\prime, u^{\prime c}$ are not needed. However, 
diagrams involving the twin EW gauge bosons introduce isospin breaking with 
one-loop size, requiring fine tuning to achieve a phenomenologically viable 
$\delta \lesssim 10^{-5}$.}

We can estimate the confinement scale for hidden QCD, $\Lambda_{\rm QCD}$,\footnote{Note that $\Lambda_{\rm QCD}$ is distinct from $\Lambda$ defined in Eq.~\eqref{eq:def_Lambda}.} by requiring~\cite{Craig:2015pha} that the visible and hidden color gauge couplings are approximately equal at the scale $\Lambda_{\rm UV} \lesssim 4\pi f$, where the theory needs to be extended. Including, in addition to the three light quark flavors, the two degenerate top partners with mass $\simeq (f/v)\, m_t /\sqrt{2}\,$, and allowing for $| (\hat{g}_s - g_s)/g_s | < 0.2$ at $\Lambda_{\rm UV} = 5$~TeV, we obtain $0.12 < \Lambda_{\rm QCD} / \mathrm{GeV} < 4.7$, where we took $f = 750$~GeV for illustration.\footnote{We used 
$2$-loop running and $\alpha_s(m_Z) = 0.118$ as input. For reference, the same running procedure applied to the SM leads to $\Lambda_{\rm QCD,\, SM} \approx 370\;\mathrm{MeV}$.} The mass scale required for viable $3\to 2$ freezeout of the dark mesons falls toward the lower end of this range.

While the Higgs portal interactions decouple early, at temperatures around a few GeV, the kinetic mixing between the hidden hypercharge gauge field $\hat{B}$ and its SM counterpart can maintain kinetic equilibrium between the two sectors until the DM freezes out. The gauge Lagrangian reads
\beq
 \mathcal{L}_g^{\rm twin} = (D_\mu \widehat{H})^\dagger D^\mu \widehat{H} - \frac{1}{4} \hat{B}_{\mu\nu}\hat{B}^{\mu\nu}  +  
 \frac{1}{2} m^2_{\hat{B}} \hat{B}_\mu \hat{B}^{\mu} + 
 \frac{\varepsilon}{2} \hat{B}^{\mu\nu} B_{\mu\nu} \,.
\eeq 
The diagonalization of the kinetic and mass Lagrangian for the four neutral gauge 
fields $\hat{W}^3, \hat{B}, W^3, B$ was performed in Ref.~\cite{Chacko:2019jgi}. 
For $m_{\hat{B}} < m_Z$ the mass of the physical dark photon is $m_{A'} \approx \hat{c}_w m_{\hat{B}}\,$, where $\hat{c}_w$ is the cosine of the 
twin weak mixing angle, and the $A'$ coupling to the SM fermions $f$ is $\varepsilon e c_w 
\hat{c}_w Q_f$. We do not specify whether $m_{\hat{B}}$ comes from a St\"uckelberg mechanism (see Ref.~\cite{Ruegg:2003ps} for a review) or from an additional dark Higgs. In either case, attention must be paid to avoid introducing a naturalness problem related to the $A'$ mass, which needs to satisfy 
$\Lambda \lesssim m_{A'} < m_Z$ for viable SIMP phenomenology.

An embedding of SIMP DM in the Twin Higgs framework was presented in Ref.~\cite{Hochberg:2018vdo}, where a complete mirror spectrum was introduced and an exact $SU(2)$ flavor symmetry relating the first two generations was imposed. The more minimal proposal sketched in Eq.~\eqref{eq:model1} aims at a more direct connection between the heavy degrees of freedom essential for Higgs naturalness and the light quark spectrum dictating DM phenomenology. We leave a detailed study of this completion to future work.

%%%%%%%%%%%%%%%%%%%%%%%%%%%%%%%%%%%%%%%%%%%%%%%%%%%%%%%%%%%%%%%%%%%%%%%%%%%%%%%
%%%%%%%%%%%%%%%%%%%%%%%%%%%%%%%%%%%%%%%%%%%%%%%%%%%%%%%%%%%%%%%%%%%%%%%%%%%%%%%%
%%%%%%%%%%%%%%%%%%%%%%%%%%%%%%%%%%%%%%%%%%%%%%%%%%%%%%%%%%%%%%%%%%%%%%%%%%%%%%%%%
\section{Dark meson decays}
\label{sec:mesondecay}
In this section we calculate the lifetimes of the hidden mesons, which provide key inputs to the analysis of cosmological and astrophysical constraints discussed below in Section~\ref{sec:constraints}. First we focus on the singlet $\eta$, the fastest-decaying meson since it is not protected by any symmetry. We then 
consider decays of the DM components: we discuss $\pi_0$ decays induced by small isospin breaking, and show that the neutral kaons are accidentally stable even when the isospin is broken by the quark masses, due to a residual $U(1)$ symmetry.

%%%%%%%%%%%%%%%%%%%%%%%%%%%%%%%%%%%%%%%%%%%%%%%%%%%%%%%%%%%%%%%%%%%%%%%%%%%%%%%%
\subsection{$\eta$ decay}
\label{subsec:etadecay}
\begin{figure}[t]
\begin{center}
\includegraphics[width=0.6\textwidth]{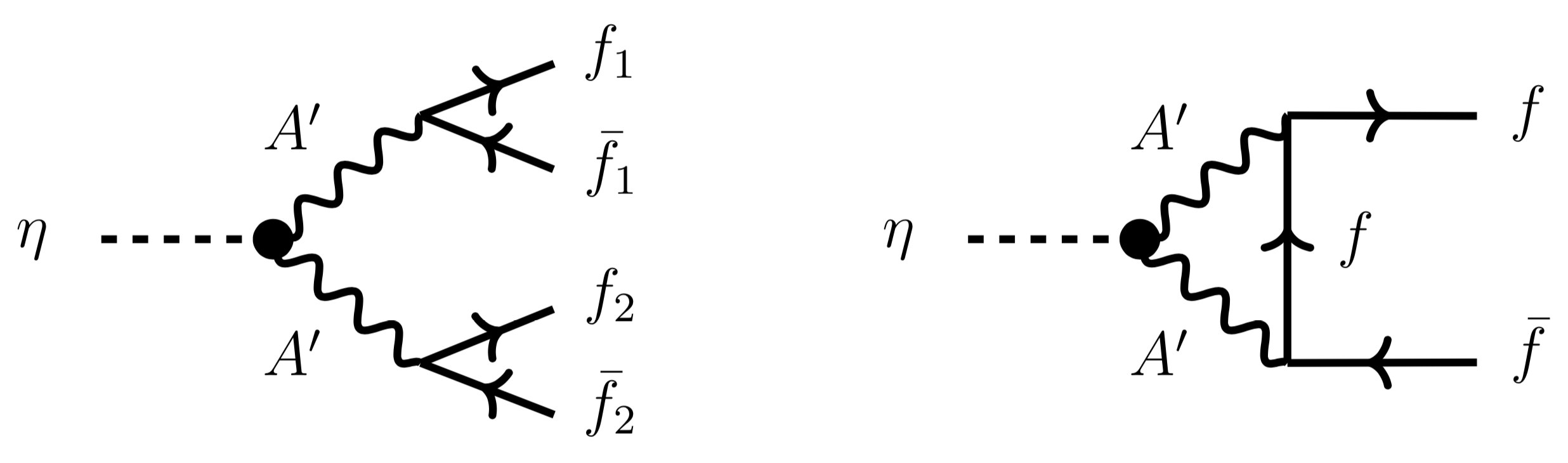}  
\caption{\label{fig:diagrams} Feynman diagrams for $\eta$ decay to SM fermions through virtual dark photon exchange.} 
\end{center}
\end{figure}

The $\eta$, being a singlet under $SU(2)_U \times U(1)_Q$, is unstable even when this is an exact symmetry. The diagrams mediating its decay are shown in Fig.~\ref{fig:diagrams}. The strength of the $\eta A^{\prime \ast}A^{\prime\ast}$ interaction, and therefore the $\eta$ lifetime, depend on the choice of hidden electric charges. 

For the standard choice of $Q$, the coupling is dominated by the AVV anomaly (see the second line of Eq.~\eqref{eq:ChPT}). For the tree-level\footnote{To avoid confusion, the counting of loops always refers to the chiral Lagrangian and not to the underlying quark-level description.} decay to four electrically-charged SM fermions, since $m_{A'} > m_\eta$, the amplitude can be matched to the effective operator
\begin{equation}
K_2 \eta\, \epsilon^{\mu\nu\rho\sigma} \partial_\sigma (\overline{f}_1 \gamma_\mu f_1) \partial_\rho (\overline{f}_2 \gamma_\nu f_2)\,,\quad\, \mathrm{with}\quad K_2^{\rm (anomaly)} = \frac{N_c\hspace{0.3mm} \varepsilon^2 \hat{e}^2 e^2 c_w^2  Q_{f_1} Q_{f_2}}{36 \pi^2 f_\pi  m_{A'}^4}\, 2 \sqrt{3}\,. \label{eq:K2_decay}
\end{equation}
Here we have taken two distinct fermion-antifermion pairs to avoid subtleties with identical particles; we will adjust the symmetry factors as appropriate when discussing the relevant case $\eta \to 4e$. We have computed the decay width corresponding to the operator in Eq.~\eqref{eq:K2_decay} with the help of FeynRules~\cite{Alloul:2013bka} and MadGraph5~\cite{Alwall:2014hca}, assuming massless fermions. The result is
\begin{equation}
\Gamma(\eta \to f_1 \bar{f}_1 f_2 \bar{f}_2) \approx \frac{(K_2)^2 m_\eta^{11}}{6301 \times 8\pi (4\pi)^4}  \approx \frac{2048}{6301} \Big( \frac{8 \sqrt{3} N_c Q_{f_1} Q_{f_2}}{9} \Big)^2 \frac{\varepsilon^4 \hat{\alpha}^2 \alpha^2 c_w^4 }{8\pi (4\pi)^4} \frac{(m_{\eta}/2)^{11}}{f_\pi^2 m_{A'}^8}\,. 
\end{equation}
Focusing on the decay to $4e$, we find the lifetime
\beq
\qquad \tau_{\eta \to 4e}^{(\rm anomaly)} \approx 1.9  \times 10^{14}\;\mathrm{s} \left( \frac{10^{-5}}{\varepsilon} \right)^4 \bigg(\frac{\alpha}{\hat{\alpha}} \bigg)^2 \bigg( \frac{m_{A'}}{0.4\;\mathrm{GeV}} \bigg)^8  \left( \frac{200\;\mathrm{MeV}}{m_\eta} \right)^9 \left( \frac{10}{m_{\eta}/f_\pi} \right)^2\,, 
\eeq
where we have included an extra factor of $2$ to approximately correct for the identical particles. For the helicity-suppressed, one-loop decay to two fermions we estimate
\begin{align} \label{eq:eta_ff_anom}
\Gamma(\eta \to f\bar{f}) &\sim  7 Q_f^4\, \frac{\varepsilon^4 \hat{\alpha}^2 \alpha^2 c_w^4 }{8\pi (4\pi)^4} \frac{(m_{\eta}/2)^5 m_f^2}{f_\pi^2 m_{A'}^4}\sqrt{1 - \frac{4m_f^2}{m_\eta^2}} \,, \\
\; \tau_{\eta \to \mu\mu}^{(\rm anomaly)} &\sim 8  \times 10^{11}\;\mathrm{s}  \left( \frac{ 10^{-5}}{\varepsilon} \right)^4 \bigg(\frac{\alpha}{\hat{\alpha}} \bigg)^2 \bigg( \frac{m_{A'}}{0.5\;\mathrm{GeV}} \bigg)^4 \left( \frac{250\;\mathrm{MeV}}{m_\eta} \right)^3 \left( \frac{10}{m_{\eta}/f_\pi} \right)^2 , \nonumber
\end{align} 
where in the second line we have focused on the decay to $\mu\mu$, which is kinematically open in part of the parameter space we consider. Note that when $m_{\eta}\gtrsim 2m_\mu$, the $ee\mu\mu$ channel is also open. 

With the $Q'$ charges, the AVV anomalies vanish but the $\eta A^{\prime \ast}A^{\prime\ast}$ interaction is still generated by higher-order, $O(p^6)$ operators in the chiral Lagrangian~\cite{Berlin:2018tvf}. Operators with one insertion of the quark mass matrix include for example
\begin{align}
&\frac{ d_1 \hat{\alpha}}{(4\pi)^2 f_\pi}\, i \epsilon^{\mu\nu\rho\sigma} \hat{F}_{\mu \nu} \hat{F}_{\rho\sigma} \mathrm{Tr}(\widehat{Q})  \mathrm{Tr}(\widehat{Q} M \Sigma^\dagger) + \mathrm{h.c.}\, , \label{eq:eta_decay_op} \\
&\frac{ d_2  \hat{\alpha}}{(4\pi)^2 f_\pi}\, i \epsilon^{\mu\nu\rho\sigma} \hat{F}_{\mu \nu} \hat{F}_{\rho\sigma}  \mathrm{Tr}(M^\dagger \Sigma \widehat{Q} \Sigma^\dagger \widehat{Q} \Sigma) + \mathrm{h.c.}\,, 
\end{align}
where $d_{1,2}$ are $O(1)$ coefficients.\footnote{Our naive dimensional analysis (NDA) estimate for the operator in Eq.~\eqref{eq:eta_decay_op} has an extra factor of $1/(4\pi)$ compared to Ref.~\cite{Berlin:2018tvf}.} Since $i \mathrm{Tr} (Q' M \Sigma^\dagger) + \mathrm{h.c.} \supset 2 (2 m + \Delta m) \eta / (\sqrt{3} f_\pi)\,$ and $i \mathrm{Tr}(M^\dagger \Sigma Q'\Sigma^\dagger Q'\Sigma) + \mathrm{h.c.} \supset - \Delta m \hspace{0.3mm} \eta / (\sqrt{3} f_\pi)$, the operator in Eq.~\eqref{eq:eta_decay_op} gives a larger contribution. By matching to Eq.~\eqref{eq:K2_decay}, we find
\begin{equation} \label{eq:K2_d1}
K_2^{(d_1)} \approx - \frac{d_1 m_\eta^2}{  (4\pi f_\pi)^2}\, K_2^{(\rm anomaly)}\,,
\end{equation}
where we have taken $B = 4\pi f_\pi$ and $N_c = 3$. In addition, there are $O(p^6)$ operators without mass insertions but containing additional derivatives, such as
\begin{equation}
\frac{d_3 \hat{\alpha}}{2 (4\pi)^3 f_\pi^2}\, i \epsilon^{\mu\nu\rho\sigma} \hat{F}_{\mu\nu} \hat{F}_{\rho\sigma} \mathrm{Tr}(\widehat{Q}) \mathrm{Tr} (\widehat{Q}\Sigma \partial_\alpha \partial^\alpha \Sigma^\dagger) + \mathrm{h.c.}\,.  
\end{equation}
This operator leads to $K_2^{(d_3)} = d_3 m_\eta^2 K_2^{(\rm anomaly)}/ (4\pi f_\pi)^2$, the same result as in Eq.~\eqref{eq:K2_d1} up to a sign. In the remainder of this paper we assume that the decays of $\eta$ are mediated by $d_1$, but it should be kept in mind that this coefficient actually represents a combination of several coefficients of $O(p^6)$ operators.

The result in Eq.~\eqref{eq:K2_d1} illustrates that the decay through higher-order operators is strongly suppressed close to the chiral limit. However, in practice we consider mesons that are only moderately lighter than $4\pi f_\pi$, and the resulting lifetime is only mildly longer than in the anomalous case,
\begin{align}
\tau_{\eta \to 4e}^{(d_1)} \approx 1.9 \times 10^{15}\;\mathrm{s} \left( \frac{10^{-5}}{\varepsilon} \right)^4 \bigg(\frac{\alpha}{\hat{\alpha}} \bigg)^2 \bigg( \frac{m_{A'}}{0.4\;\mathrm{GeV}} \bigg)^8  \left( \frac{200\;\mathrm{MeV}}{m_\eta} \right)^9 \left( \frac{10}{m_{\eta}/f_\pi} \right)^6 \Big( \frac{0.5}{d_1}\Big)^2 .
\end{align}
For the two-body decay to $\mu\mu$, our estimate is
\begin{equation} \label{eq:eta_ff_nonanom}
\tau_{\eta \to \mu\mu}^{(d_1)} \sim 8 \times 10^{12}\;\mathrm{s}  \left( \frac{ 10^{-5}}{\varepsilon} \right)^4 \bigg(\frac{\alpha}{\hat{\alpha}} \bigg)^2 \bigg( \frac{m_{A'}}{0.5\;\mathrm{GeV}} \bigg)^4 \left( \frac{250\;\mathrm{MeV}}{m_\eta} \right)^3 \left( \frac{10}{m_{\eta}/f_\pi} \right)^6 \Big( \frac{0.5}{d_1}\Big)^2  .
\end{equation} 
Note that the above amplitudes do not mediate the decay of $\eta$ to SM $\pi_0 \ell^{+} \ell^{-}$, as axial-vector couplings are not involved. The results for the decay to $f\bar{f}$ in Eqs.~\eqref{eq:eta_ff_anom} and \eqref{eq:eta_ff_nonanom} only represent rough estimates and should therefore be taken with some caution. 

%%%%%%%%%%%%%%%%%%%%%%%%%%%%%%%%%%%%%%%%%%%%%%%%%%%%%%%%%%%%%%%%%%%%%%%%%%%%%%%%%%
\subsection{Dark matter decay}
\label{subsec:dmdecays}
The neutral pion $\pi_0$ decays through its mixing with $\eta$, which is proportional to the isospin-breaking parameter $\delta$ defined in Eq.~\eqref{eq:delta_param}. This leads parametrically to $\tau_{\pi_0} \sim \delta^{-2} \tau_{\eta}\hspace{0.2mm}$. 

In the anomalous case we find that the amplitude for $\pi_0 \to 4f$ is given by Eq.~\eqref{eq:K2_decay} with the replacement $2\sqrt{3} \to -\, 3\hspace{0.3mm} \delta /2\,$, yielding
\begin{equation} \label{eq:pi0_4e_anom}
\tau_{\pi_0 \to 4e}^{\rm (anomaly)} \sim 1.0  \times 10^{25}\;\mathrm{s}  \left( \frac{10^{-5}}{\delta} \right)^2 \left( \frac{10^{-5}}{\varepsilon} \right)^4 \bigg(\frac{\alpha}{\hat{\alpha}} \bigg)^2 \bigg( \frac{m_{A'}}{0.4\;\mathrm{GeV}} \bigg)^8  \left( \frac{200\;\mathrm{MeV}}{m_{\pi_0}} \right)^9 \left( \frac{10}{m_{\pi_0}/f_\pi} \right)^2 \hspace{-1mm} . \hspace{-1mm}
\end{equation}
For the one-loop helicity-suppressed decay to $\mu\mu$ we similarly find
\begin{equation} \label{eq:pi0_mumu_anom}
\tau_{\pi_0 \to \mu\mu}^{\rm (anomaly)} \sim 4.1  \times 10^{22}\;\mathrm{s} \left( \frac{10^{-5}}{\delta} \right)^2 \left( \frac{ 10^{-5}}{\varepsilon} \right)^4 \bigg(\frac{\alpha}{\hat{\alpha}} \bigg)^2 \bigg( \frac{m_{A'}}{0.5\;\mathrm{GeV}} \bigg)^4 \left( \frac{250\;\mathrm{MeV}}{m_{\pi_0}} \right)^3 \left( \frac{10}{m_{\pi_0}/f_\pi} \right)^2 \hspace{-1mm}.
\end{equation} 

In the scenario without AVV anomalies we obtain $i \mathrm{Tr} (Q' M \Sigma^\dagger) + \mathrm{h.c.} \supset - m\hspace{0.2mm} \delta\hspace{0.2mm} \pi_0/ f_\pi\,$, and from Eq.~\eqref{eq:eta_decay_op} we derive for $\pi_0\to 4e$
\begin{equation}  \label{eq:pi0_4e_nonanom}
\tau_{\pi_0 \to 4e}^{(d_1)} \sim 1.0 \times 10^{26}\;\mathrm{s} \left( \frac{10^{-5}}{\delta} \right)^{\hspace{-0.5mm}2} \hspace{-1mm} \left( \frac{10^{-5}}{\varepsilon} \right)^4 \hspace{-1mm}\bigg(\frac{\alpha}{\hat{\alpha}} \bigg)^{\hspace{-0.5mm}2} \hspace{-0.75mm} \bigg( \frac{m_{A'}}{0.4\;\mathrm{GeV}} \bigg)^{\hspace{-0.5mm}8}\hspace{-1mm}  \left( \frac{200\;\mathrm{MeV}}{m_{\pi_0}} \right)^{\hspace{-0.5mm}9} \hspace{-1mm} \left( \frac{10}{m_{\pi_0}/f_\pi} \right)^6 \hspace{-1mm} \Big( \frac{0.5}{d_1}\Big)^2 \hspace{-1mm},\hspace{-1mm}
\end{equation}
taking $B = 4 \pi f_\pi$. Our estimate for the decay to $\mu\mu$ is
\begin{align} \label{eq:pi0_ff_nonanom}
\tau_{\eta \to \mu\mu}^{(d_1)} \sim 4.1 \times 10^{23}\;\mathrm{s} \left( \frac{10^{-5}}{\delta} \right)^{\hspace{-0.5mm}2} \left( \frac{ 10^{-5}}{\varepsilon} \right)^4 & \bigg(\frac{\alpha}{\hat{\alpha}} \bigg)^2 \bigg( \frac{m_{A'}}{0.5\;\mathrm{GeV}} \bigg)^4  \\ 
&\qquad \times \left( \frac{250\;\mathrm{MeV}}{m_{\pi_0}} \right)^3 \left( \frac{10}{m_{\pi_0}/f_\pi} \right)^6 \Big( \frac{0.5}{d_1}\Big)^2  . \nonumber
\end{align} 
While detailed constraints on DM decay are discussed in Section~\ref{sec:DM_decays}, we anticipate that the $\pi_0$ lifetime must be longer than about $10^{25}$ seconds to be phenomenologically viable.

For the neutral kaon $K_0$, we note that the explicit isospin breaking caused by $dm$ preserves a residual $U(1)$ symmetry under which the quarks have charges $Q_r = \mathrm{diag}\,(0, 1, -1)$, since $Q_r$ and the quark mass matrix $M$ commute. Under this symmetry the complex mesons have charges $Q_{r} (\pi_+) = -1$, $Q_{r} (K_+) = +1$, and $Q_{r} (K_0) = +2\,$. It follows that $K_0$, being the lightest particle charged under $U(1)_{Q_r}$, is accidentally stable.

%%%%%%%%%%%%%%%%%%%%%%%%%%%%%%%%%%%%%%%%%%%%%%%%%%%%%%%%%%%%%%%%%%%%%%%%%%%%%%%%%%%%
%%%%%%%%%%%%%%%%%%%%%%%%%%%%%%%%%%%%%%%%%%%%%%%%%%%%%%%%%%%%%%%%%%%%%%%%%%%%%%%%%%%%
%%%%%%%%%%%%%%%%%%%%%%%%%%%%%%%%%%%%%%%%%%%%%%%%%%%%%%%%%%%%%%%%%%%%%%%%%%%%%%%%%%%%
\section{Cosmological history} 
\label{sec:cosmo_history}
In this section we discuss various aspects of the cosmological history of our setup. We assume that the hidden and SM sectors are in equilibrium in the early Universe (the necessary conditions will be detailed below) and study the evolution and 
freezeout of the dark meson abundances. Where relevant, for convenience we use the 
labels $L, M, H$ to refer to the light ($\pi_0, K_0, \overline{K}_0$), middle
($\pi_\pm,\, K_\pm$), and heavy ($\eta$) meson states, respectively, and 
$m_{L,M,H}$ to indicate their masses.

%%%%%%%%%%%%%%%%%%%%%%%%%%%%%%%%%%%%%%%%%%%%%%%%%%%%%%%%%%%%%%%%%%%%%%%%%%%%%%%%%%%
\subsection{Freezeout and relic abundances}
\label{sec:relic_abundance}
We begin by discussing the thermal freezeout and relic abundances of the dark mesons. As is well known, the $3\to 2$ strong interaction processes mediated by the WZW term can remain efficient in keeping dark mesons in chemical equilibrium down to $x \equiv m_{\pi_0}/T\approx 20$, thereby setting the relic density of the lightest multiplet to the observed value $Y_{\rm DM} \approx 4.1\times 10^{-10}\, (\mathrm{GeV} / m_{\rm DM})$, where $Y \equiv n/s$ is the number density normalized to the total entropy. We will see that $2\to 2$ scattering processes among dark mesons remain active down to much lower temperatures, strongly suppressing the abundances of the heavier multiplets, in particular of the unstable $\eta$. The complete Boltzmann equations that we solve to obtain the evolution of the meson abundances in the early Universe are provided
in Appendix~\ref{sec:BEs}.

We study the evolution of the system with the following leading-order spectrum,
\begin{equation} \label{eq:LO_spectrum}
m_{\pi_0, K_0, \overline{K}_0} = 200\;\mathrm{MeV}, \qquad m_{\pi_\pm, K_\pm} = 210\;\mathrm{MeV}, \qquad m_{\eta} = 213\;\mathrm{MeV},
\end{equation}
obtained by taking $B = 4\pi f_\pi = 265$~MeV, $m = 76$~MeV, and $\Delta m = 15$~MeV. The relative splittings with respect to the lightest multiplet are $\Delta_{\pi} = 0.05$ and $\Delta_{\eta} \simeq 4\Delta_\pi/3 = 0.066$, chosen to be comparable to $T_{\rm fo}^{3\to 2}/m_{\pi_0} \approx 1/20$. Note that we have $2 m_{\pi_+} > m_{\pi_0} + m_\eta$, so that $\pi_+\pi_-\to\pi_0\hspace{0.3mm}\eta$ scattering is kinematically open at zero temperature.

\begin{figure}[!t]
\begin{center}
\includegraphics[width=0.485\textwidth]{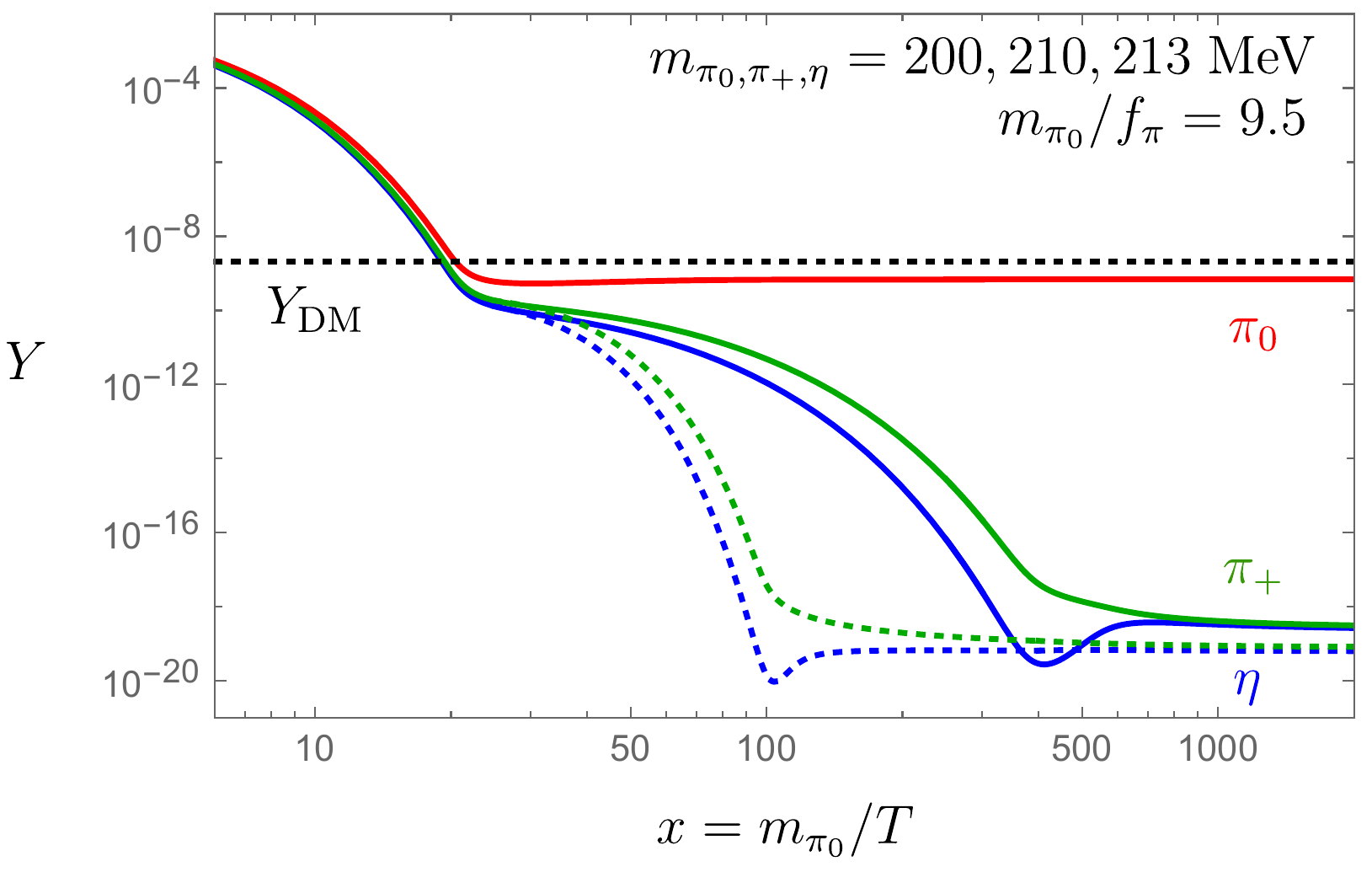}  \hspace{1mm}
\includegraphics[width=0.485\textwidth]{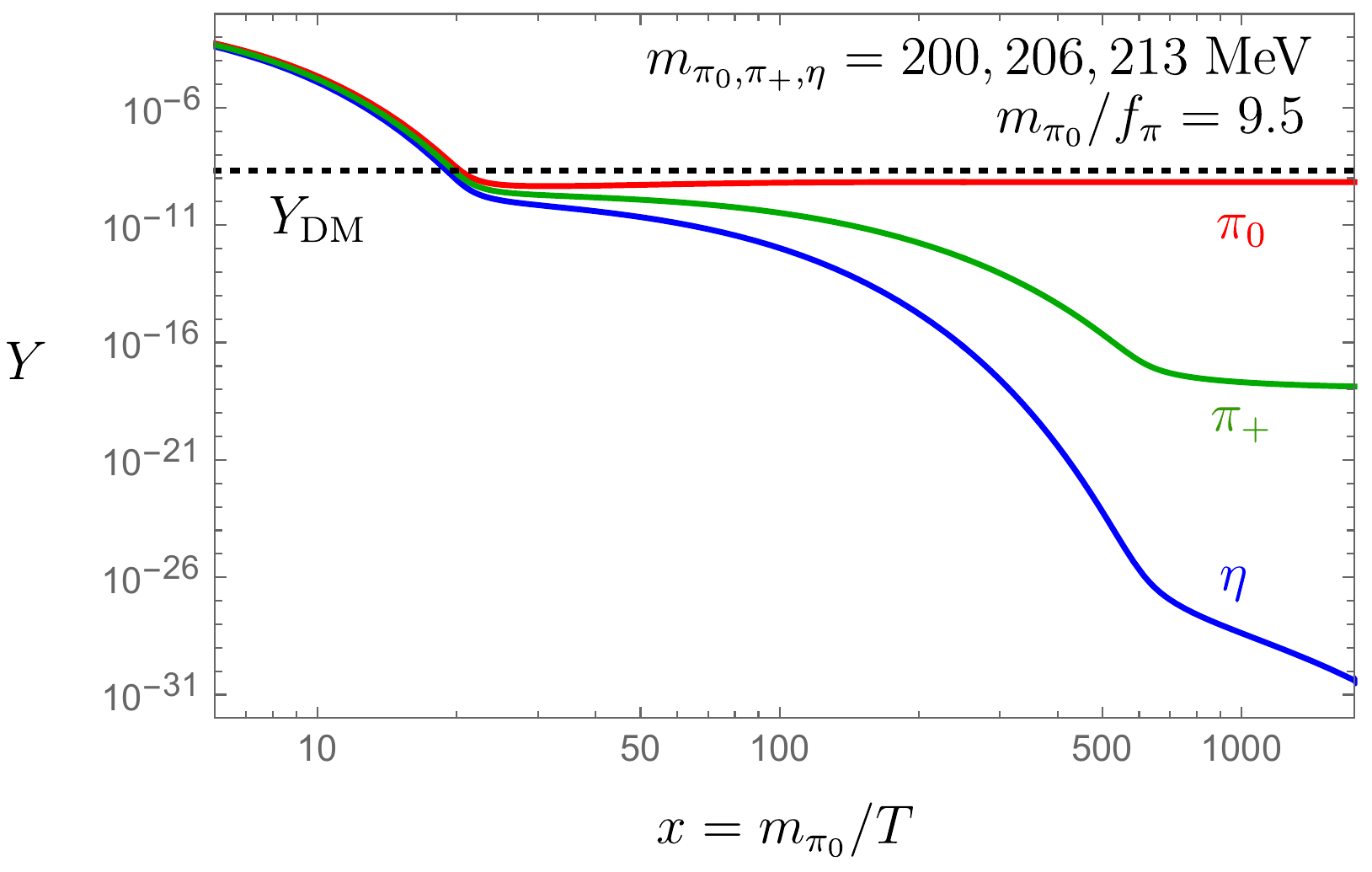} 
%\vspace{4mm}
\caption{\label{fig:Boltzmann} {\it Left panel:} Boltzmann evolution for the leading-order spectrum of Eq.~\eqref{eq:LO_spectrum} satisfying $2 m_{\pi_+} > m_{\pi_0} + m_\eta$. Solid curves assume kinetic coupling between the SM and hidden sectors throughout, whereas dotted curves correspond to decoupling at $x_{\rm dec} = 25$. Note that each curve denotes the abundance of a single real degree of freedom; in particular, the total DM abundance is given by $3 Y_{\pi_0}$ and matches the observed value (indicated by the black dotted line). {\it Right panel:} Illustration of an alternative scenario with a relatively larger $m_\eta$, satisfying $2 m_{\pi_+} < m_{\pi_0} + m_\eta$, resulting in strong further depletion of the $\eta$ abundance.}
\end{center}
\end{figure}

The evolution of the meson abundances for these parameters is shown in the left panel of Fig.~\ref{fig:Boltzmann}. Early on, for $m_{\pi_0}/T<20$, number-changing $3\to2$ processes and dark meson-SM scatterings keep all dark mesons in chemical and kinetic equilibrium with the SM bath, so that their abundances follow the Boltzmann-suppressed thermal distributions with $Y_{\eta} < Y_{\pi_+} < Y_{\pi_0}$. The 5-point interactions enable every meson to scatter with two states belonging to the lightest multiplet, which are the most abundant of the dark mesons; hence all $3\to 2$ processes freeze out roughly when
\begin{equation} \label{eq:sigmav2_def}
(n^{\rm eq}_{\pi_0})^2 \hspace{0.2mm} \frac{1}{6}\hspace{0.2mm} \langle \sigma v^2 \rangle_0 \sim H \,, \qquad \langle \sigma v^2 \rangle_0 \equiv \frac{(5!)^2 m_{\pi_0}^5 N_c^2}{96\sqrt{5}\, \pi^5 2^{10} f_\pi^{10} x^2}\,.
\end{equation}
As is characteristic of the SIMP mechanism, this freezeout occurs roughly at $ x \approx 20$, after which the abundances of $\pi_0,K_0,\overline{K}_0$ remain approximately constant.

The abundances of the heavier mesons continue to deplete even after the $3\to2$ processes freeze out, thanks to $2\to 2$ annihilation processes such as $HH\to MM, MM\to LL, MM\to HL$, which remain active.\footnote{As an aside, we note that the $HLMM$ interactions combined with kinetic mixing cannot mediate decays of the DM triplet to SM particles, due to symmetry arguments.} These interactions maintain the densities of the heavier mesons on the ``shifted'' equilibrium curves $Y_i^{\rm eq} (Y_{\pi_0} / Y_{\pi_0}^{\rm eq}) \simeq Y_{\pi_0} e^{- \Delta_i x}$~\cite{Hochberg:2018vdo}. Note that these annihilations provide negligible corrections to the frozen-out abundances of the $L$ states, as $Y_\eta,Y_{\pi_+} \ll Y_{\pi_0}$ at this stage. Assuming kinetic coupling between the hidden and SM sector, our numerical results show that the $2\to 2$ annihilations begin to freeze out around $x \approx 400$ (see solid curves in the left panel of Fig.~\ref{fig:Boltzmann}). To estimate the freezeout abundance $Y_{\pi_+}$, we can apply the instantaneous freezeout approximation to the $MM\to LL$ process as
\begin{equation} \label{eq:Mfreezeout}
e^{ - \Delta_\pi x} n_{\pi_0} 3\hspace{0.2mm} \langle \sigma v \rangle_{MM\to LL} \sim H \,, \qquad Y_{\pi_+}^{\rm fo} \approx Y_{\pi_0} \, e^{-\Delta_\pi x_+^{\rm fo}}\,,
\end{equation}
where the thermally-averaged $2\to 2$ cross section is defined in Eq.~\eqref{eq:2to2_sigmav}. For our parameters, this gives $x_{+}^{\rm fo} \approx 420$ and $Y_{\pi_+}^{\rm fo} \approx 4.5 \times 10^{-19}$, in good agreement with the numerical results. 

The evolution of the $\eta$ abundance is more complex. After $3\to2$ freezeout, $Y_\eta$ decreases due to the $HH\to MM, HH\to LL$ processes, as well as the thermally driven $HL\to MM$. Around $x \approx 400$, all of these processes become inefficient at depleting $\eta$; as a consequence, $Y_\eta$ departs from the shifted equilibrium curve and undergoes a short period of {\it increase} due to injections from the $MM \to HL$ processes, which are still active. The increased $\eta$ density eventually leads to a re-coupling of $HH\to MM$, and the final ratio of the $\eta$ and $\pi_+$ abundances is determined by detailed balance between these two processes,
\begin{equation}
\frac{Y_\eta^{\rm fo}}{Y_{\pi_+}^{\rm fo}} \simeq \sqrt{\frac{36\, \langle \sigma v \rangle_{MM\to HL}}{49\, \langle \sigma v \rangle_{HH\to MM} }} \,.
\end{equation}
For the spectrum under consideration, this ratio is $\approx 0.89$. The key to understanding this behavior is to observe that both $MM\to HL$ and $HH\to MM$ are kinematically allowed at $T=0\,$; hence the freezeout abundance is driven not by the familiar Boltzmann suppression, but by detailed balance between different processes. Furthermore, note that since $Y_\eta \ll Y_{\pi_+}$ at $x_{+}^{\rm fo}$ (when the $M$ states freeze out), such interplay only gives a mild correction to $Y_{\pi_+}^{\rm fo}$.

The above estimates assume that the SM and dark sectors remain in kinetic equilibrium throughout, via the scattering of $\pi_\pm, K_\pm$ on electrons. However, depending on the value of $\hat{y} \equiv \varepsilon \hat{e} (m_{\pi_0}/m_{A'})^2$, the two sectors may decouple at some point in the evolution. After decoupling, the hidden sector redshifts like matter and thus its temperature decreases faster, $T_D \propto a^{-2}$ (with $a$ the scale factor), compared to the SM bath temperature, which decreases as $T \propto a^{-1}$, giving a ratio $T_D / T \simeq x_{\rm dec}/x$ for $x > x_{\rm dec}$. To take this effect into account, Eq.~\eqref{eq:Mfreezeout} can be modified as
\begin{equation}
e^{ - \Delta_\pi x^2/x_{\rm dec}}\, n_{\pi_0} 3\hspace{0.2mm} \langle \sigma v \rangle_{MM\to LL} \sim H \,, \qquad Y_{\pi_+}^{\rm fo} \approx Y_{\pi_0} \, e^{-\Delta_\pi (x_+^{\rm fo})^2/ x_{\rm dec}}\, .
\end{equation}
Assuming $x_{\rm dec} = 25$, i.e. the two sectors decouple immediately after the light states $L$ freeze out, results in $x_+^{\rm fo} \approx 110$ (corresponding to $(x_+^{\rm fo})^2 / x_{\rm dec} \approx 450$) and $Y_+^{\rm fo} \approx 1.1 
\times 10^{-19}$, a factor $\sim 4$ smaller than in the coupled scenario. These estimates are borne out in the numerical analysis, see the dotted curves in the left panel of Fig.~\ref{fig:Boltzmann}. Earlier decoupling of the two sectors therefore results in further $O(1)$ suppression of the $M$ and $H$ abundances.

In the right panel of Fig.~\ref{fig:Boltzmann}, we consider a scenario where $MM \to HL$ scattering is kinematically closed at $T=0$. This can occur if higher-order effects raise the $\eta$ mass above $(2m_{\pi_+} - m_{\pi_0})$. Corrections to the masses of the pNGBs arise from $O(p^4)$ operators with two insertions of the quark mass matrix $M$~\cite{Gasser:1984gg}, among which
\begin{equation} \label{eq:etamass_NLO}
\frac{c_7 B^2}{(4\pi)^2}  \big[ \mathrm{Tr} (M\Sigma^\dagger - \Sigma M^\dagger)\big]^2 \,\supset -\, \frac{16\, c_7 B^2}{3 (4\pi)^2 f_\pi^2} \frac{}{} (\Delta m)^2 \eta^2\,
\end{equation}
only affects the $\eta$ mass~\cite{Berlin:2018tvf}. The NDA size of the coefficient is $c_7 \sim O(1)$. In the SM, its sign is negative due to $\eta\,$--$\,\eta'$ mixing. In our framework, however, since we require $m_{\pi}/f_\pi \sim 8\,$--$\,10$, which is 
much larger than in the SM, we cannot a priori exclude the scenario where $\eta'$ would be {\it lighter} than $\eta$, resulting in $c_7 > 0$ and thus raising the $\eta$ mass. In the right panel of Fig.~\ref{fig:Boltzmann} we illustrate this scenario, assuming the input parameters $B = 265$~MeV, $m = 76$~MeV, $\Delta m = 9$~MeV, and $c_7 = +\hspace{0.15mm} 2$. The $\eta$ abundance continues to be efficiently depleted by the $HL\to MM$ processes even at very low temperatures, resulting in an enormous suppression. While this alternative possibility deserves to be kept in mind, in the light of the above discussion it appears somewhat less likely on theoretical grounds. Therefore, we neglect it in the rest of the discussion, focusing solely on the leading-order meson spectrum.

%%%%%%%%%%%%%%%%%%%%%%%%%%%%%%%%%%%%%%%%%%%%%%%%%%%%%%%%%%%%%%%%%%%%%%%%%%%%%%%%%%%%%
\subsection{Consequences of $\pi_0\,$--$\,K_0$ mass splitting} 
\label{sec:pi0_K0_splitting}
In the previous subsection, the small isospin breaking introduced by $dm$ could be safely neglected. At later times, however, this effect can play a role in determining the relic abundances of the DM components through $K_0 \overline{K}_0 \leftrightarrow \pi_0 \pi_0$ processes. According to Eq.~\eqref{eq:meson_masses}, isospin breaking generates a small splitting between the $\pi_0$ and $K_0$ masses,
\begin{equation} \label{eq:DeltaK}
\Delta_K \equiv \frac{m_{K_0} - m_{\pi_0}}{m_{\pi_0}} \simeq \frac{(dm)^2}{16m \Delta m} \simeq \frac{\Delta_\pi}{4}\,\delta^2 > 0\,.
\end{equation}
For $\delta \lesssim 10^{-5}$, necessary to avoid observational bounds on DM decays (see Section~\ref{sec:DM_decays}), this splitting is $\lesssim O(10^{-4})$~eV. The splitting enables $K_0 \overline{K}_0\to \pi_0 \pi_0$ annihilation to slowly convert part of the kaon population into $\pi_0$'s. We expect this process to freeze out when
\begin{equation}
e^{- \Delta_K\, x^2 /x_{\rm dec} }\, n_{\pi_0} \langle \sigma v \rangle_{K_0 \overline{K}_0 \to \pi_0 \pi_0} \sim H \,, \qquad \langle \sigma v \rangle_{K_0 \overline{K}_0 \to \pi_0 \pi_0} \simeq \frac{9\hspace{0.15mm} m_{\pi_0}^2}{64\pi f_\pi^4} \sqrt{1 - \frac{m_{\pi_0}^2}{m^2_{K_0}}}\,,
\end{equation}
where we have made the assumption that this freezeout occurs after kinetic decoupling of the two sectors. The final $K_0$ abundance is $Y_{K_0}^{\rm fo} = Y_{\pi_0} e^{- \Delta_K\, x_{\rm fo}^2 / x_{\rm dec}}$. For $\delta = 10^{-5}$ and $x_{\rm dec} = 20\,$--$\,200$, we find that freezeout occurs at $T_{D}^{\rm fo} \sim 2\,$--$\,6 \times 10^{-4}$~eV (corresponding to $T^{\rm fo} \sim 50$\,--$\,20$~eV), to be compared with the mass splitting of $2.5\times 10^{-4}$~eV, and the kaon abundance is moderately depleted to $Y_{K_0}^{\rm fo} / Y_{\pi_0} \sim 0.3\,$--$\,0.6$. For smaller $\delta \lesssim 10^{-6}$, the freezeout temperature in the dark sector is much larger than the mass splitting, and the kaon density is not appreciably suppressed. 

This mild depletion of the kaon abundance can, however, be reversed in the late Universe, when kaons are regenerated via $\pi_0 \pi_0 \to K_0 \overline{K}_0$ up-scatterings in DM halos. The regenerated fractional density can be roughly estimated as~\cite{Batell:2009vb}
\begin{equation} \label{eq:upscattering}
\frac{n_{K_0} + n_{\overline{K}_0}}{n_{K_0} + n_{\overline{K}_0} + n_{\pi_0}} \sim  \tau_{\rm int}\, \frac{\rho_{\rm DM}}{m_{\pi_0}}\, \langle \sigma v \rangle_{\pi_0 \pi_0 \to K_0 \overline{K}_0}\,,
\end{equation}
where $\tau_{\rm int}$ is the ``integration time'' over which scatterings occur. After relating the forward and backward reactions and assuming $\Delta_K \ll v^2$, the right-hand side of Eq.~\eqref{eq:upscattering} becomes
\begin{equation}
\sim  \tau_{\rm int} \frac{\rho_{\rm DM}}{m_{\pi_0}}  \frac{ 9\hspace{0.15mm} m_{\pi_0}^2}{64 \pi f_\pi^4 }\,v \sim \\ 
\frac{2}{3} \left( \frac{\tau_{\rm int}}{2\, \mathrm{Gyr}} \right) \left( \frac{\rho_{\rm DM}}{10^{-26}\, \frac{\mathrm{g}}{ \mathrm{cm}^3}} \right) \left( \frac{200\;\mathrm{MeV}}{m_{\pi_0}} \right)^3 \left( \frac{m_{\pi_0}/f_\pi}{10} \right)^4 \bigg( \frac{v}{1000\; \frac{\mathrm{km}}{\mathrm{s}}} \bigg),
\end{equation}
where we have taken as reference the DM density and velocity dispersion relevant for galaxy clusters. Therefore, kaons and pions are expected to become equally distributed within $\sim 2$ Gyr. While this is only a rough estimate, it suggests that it is appropriate to assume that DM is composed equally of $\pi_0,\,K_0,$ and $\overline{K}_0$ for late-Universe phenomena such as cluster mergers. Similar considerations apply to galaxies, where the typical velocity dispersion is $v \sim 250$ km/s, and possibly even to dwarf galaxies, where the average velocity can be as low as $30$~km/s but the DM density is larger.

The masses of $\pi_\pm$ and $K_\pm$ are also split by isospin breaking, but at $O(\delta)$, as can be read in Eq.~\eqref{eq:meson_masses}. However, since the densities of the charged mesons are 
extremely suppressed, this effect does not have appreciable phenomenological consequences, and we neglect it.

%%%%%%%%%%%%%%%%%%%%%%%%%%%%%%%%%%%%%%%%%%%%%%%%%%%%%%%%%%%%%%%%%%%%%%%%%%%%%%%%%%%
\subsection{Larger mass splittings}
We now comment on scenarios where the mass splitting between the lightest ($L$) and next-to-lightest ($M$) multiplets becomes appreciable. We will show that as long as \mbox{$3 m_L > 2 m_M$}, the SIMP mechanism proceeds largely as in the degenerate case. We first provide simple analytical arguments supporting this conclusion, followed by detailed numerical results.

Recall that the $L$ freezeout abundance is determined from detailed balance between the $3\to 2$ process $LLL\to MM$ and its inverse $2\to 3$ process $MM\to LLL$. For comparable masses $m_L\approx m_M$, the inverse process needs to make up energy equivalent to the mass of a single particle, incurring a corresponding Boltzmann suppression $e^{-m_L/T}$ from the thermal tail. In other words, the rate of the inverse process is proportional to $Y_M^2 e^{-m_L/T} 
\approx Y_L^2 e^{-m_L/T}$, whereas the rate of the forward process is proportional to $Y_L^3$, hence detailed balance between the two gives the familiar Boltzmann-suppressed abundance $Y_L\sim e^{-m_L/T}$.

When the mass splitting becomes larger than the bath temperature $T$ but $LLL\to MM$ remains open without kinematic suppression, i.e. $3 m_L > 2 m_M$, the inverse process $MM\to LLL$ needs to make up a smaller amount of energy, $3 m_L - 2 m_M$, and the incurred thermal suppression $e^{-(3 m_L - 2 m_M)/T}$ appears much weaker. However, note that the $M$ abundance itself is further suppressed in this case, $Y_M\sim e^{-m_M/T}$, hence the rate of the inverse process is proportional to $Y_M^2 e^{-(3 m_L - 2 m_M)/T}= e^{-3m_L/T}$, which is the same as in the degenerate scenario. Freezeout therefore largely proceeds as in the degenerate case, resulting in the same freezeout temperature $T_{\rm fo}^{3\to 2}$, and the SIMP mechanism still produces the correct relic density. 

\begin{figure}[t!]
\begin{center}
\includegraphics[width=0.49\textwidth]{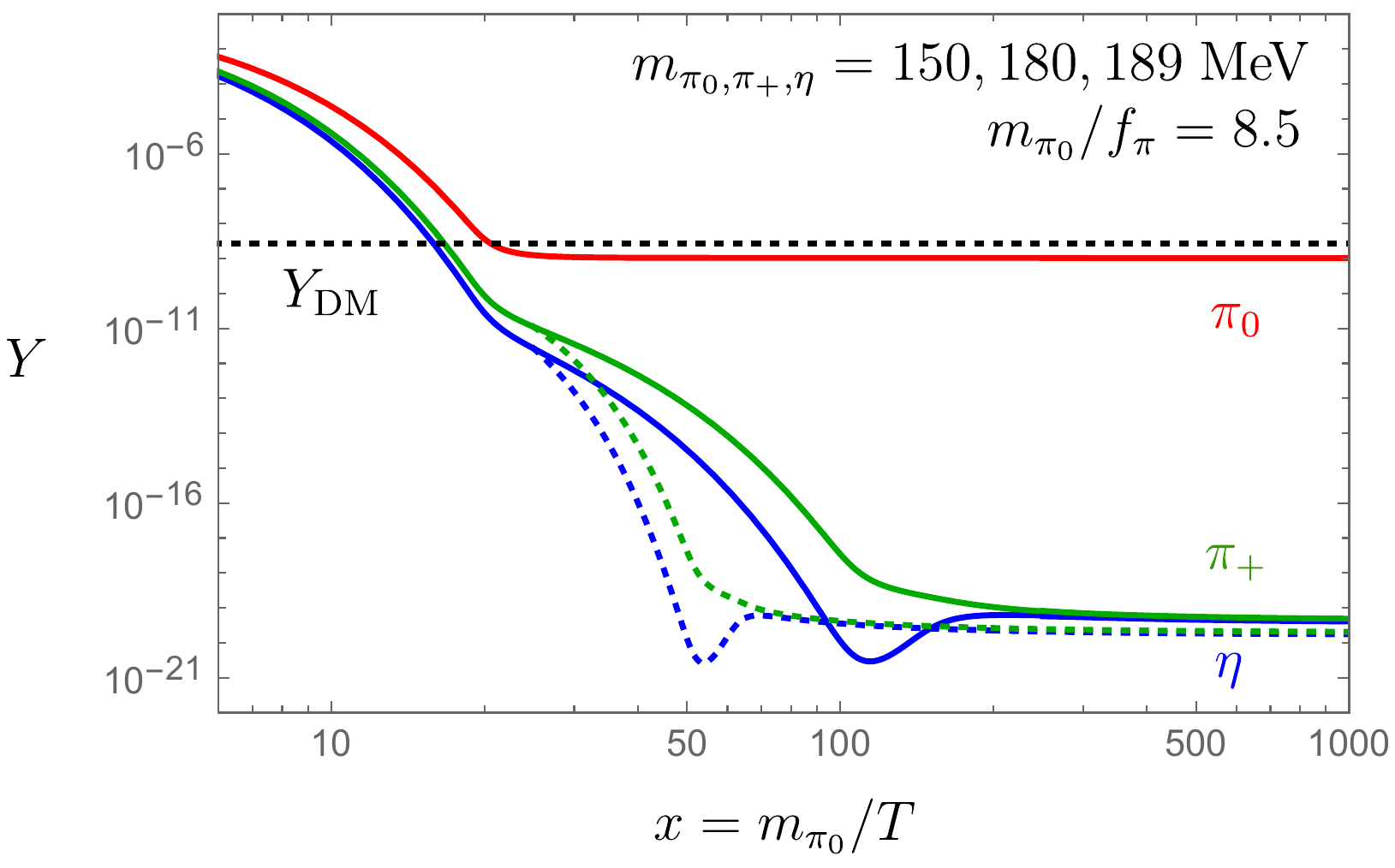} 
\caption{\label{fig:Boltzmann_2} Boltzmann evolution for the spectrum in Eq.~\eqref{eq:LO_spectrum2}, characterized by meson mass splittings larger than the $3\to 2$ freezeout temperature. Solid curves assume kinetic coupling between the SM and hidden sector, whereas dotted curves correspond to decoupling at $x_{\rm dec} = 25$.} 
\end{center}
\end{figure}

We verify these findings numerically with the following mass spectrum,
\begin{equation} \label{eq:LO_spectrum2}
m_{\pi_0, K_0, \overline{K}_0} = 150\;\mathrm{MeV}, \qquad m_{\pi_\pm, K_\pm} = 180\;\mathrm{MeV}, \qquad m_{\eta} = 189\;\mathrm{MeV},
\end{equation}
obtained by taking $B = 4\pi f_\pi = 222$~MeV, $m = 51$~MeV, and $\Delta m = 45$~MeV. The relative splittings with respect to the lightest multiplet are $\Delta_{\pi} = 0.20$ and $\Delta_{\eta} \simeq 4\Delta_\pi/3 = 0.26$, much larger than $T_{\rm fo}^{3\to 2}/m_{\pi_0}\approx 0.05$. We emphasize that the quark masses in this case roughly satisfy $m_u \sim 2\hspace{0.15mm} m_{d,s}$, namely, there is an $O(1)$ mass splitting between the up and the down-type quarks. The results are illustrated in Fig.~\ref{fig:Boltzmann_2}. For simplicity, in our numerical analysis we continue to employ the expression for $\langle \sigma v^2\rangle_{0}$ computed in the degenerate limit; the corrections to this approximation are expected to be subleading compared to the Boltzmann suppressions from the meson mass splittings. The constraints from $\eta$ decay will be discussed later in Section~\ref{sec:eta_decays}, see Fig.~\ref{fig:etadecay}.

Incidentally, we notice that for large mass splittings the scenario where the $MM \to HL$ process is kinematically closed at $T=0$ may become more plausible, as the $O(1)$ breaking of $SU(3)$ by $\Delta m \sim m$ lifts the $\eta$ mass closer to the strong scale $\Lambda$. This in turn makes it more likely that $m_\eta > m_{\eta'}$, resulting in $c_7 > 0$ in Eq.~\eqref{eq:etamass_NLO}. In this case, the evolution of the relic abundances would be qualitatively similar to the one in the right panel of Fig.~\ref{fig:Boltzmann}.  

On the other hand, the outlook changes rapidly when $2 m_M >3 m_L$. The leading processes depleting the DM abundance in this case are $LLL\to MM$ with thermal suppression and $LLM\to LM$. The (forward) rate for the former process is proportional to $Y_L^3 e^{-(2m_M-3m_L)/T}$, and for the latter process goes as $Y_L^2 Y_M\approx Y_L^3 e^{-(m_M-m_L)/T}$ (compared to $Y_L^3$ in the degenerate case). It is then clear that the rates for these processes drop below the Hubble rate much earlier; depending on whether $LLL\to MM$ or $LLM\to LM$ is dominant, the new freezeout temperature can be estimated as
\begin{equation}
\frac{T_{\rm fo}^{3\to 2}(\text{new})}{T_{\rm fo}^{3\to 2}(\text{old})}\approx\frac{1}{2} \big[1+\Delta_\pi + \text{min}(\Delta_\pi , 1)\big]\,,
\end{equation}
valid for $\Delta_\pi > 0.5$. In this regime, we therefore expect freezeout 
to occur much earlier, leading to a DM relic abundance several orders of magnitude larger than the desired value, signaling a breakdown of the SIMP mechanism as a viable method to produce thermal DM.

%%%%%%%%%%%%%%%%%%%%%%%%%%%%%%%%%%%%%%%%%%%%%%%%%%%%%%%%%%%%%%%%%%%%%%%%%%%%%%
\subsection{Thermalization of hidden and SM sectors}
\label{sub:thermalization}
Viable SIMP freezeout requires the dark photon $A'$ to maintain kinetic equilibrium between the hidden and SM sectors down to $T_{\rm fo}^{3\to2} \simeq m_{\pi}/20$. For $m_\pi \lesssim \mathrm{GeV}$, this corresponds to $T^{3\to2}_{\rm fo} \lesssim 50$~MeV, and it is a reasonable approximation to focus on dark meson scattering with electrons and neutrinos, as muons and pions are already somewhat non-relativistic~\cite{Lee:2015gsa,Hochberg:2015vrg}. The cross section for, e.g., $\pi_+ f \to \pi_+ f$ scattering, where $f$ is a Dirac SM fermion, is mediated by the exchange of neutral vector bosons $\widetilde{V}_{A,B} = A', Z$ in the $t$-channel. Neglecting $m_f$, we find
\begin{equation}
\sigma v(\pi_+ f \to \pi_+ f) = \frac{p_f^2}{2\pi } \sum_{\widetilde{V}_A, \widetilde{V}_B} \frac{g_{\widetilde{V}_A \pi \pi } g^\ast_{\widetilde{V}_B\pi \pi } }{ m^2_{\widetilde{V}_A} m^2_{\widetilde{V}_B} } (g_{\widetilde{V}_A f \bar{f}}^V \,g^{V \ast}_{\widetilde{V}_B f \bar{f}} + g_{\widetilde{V}_A f \bar{f}}^A \,g^{A \ast}_{\widetilde{V}_B f \bar{f}})\,,
\end{equation}
where $p_f$ is the fermion three-momentum and we have averaged over the $f$ spin states. The dominant contribution comes from the exchange of $A'$, which couples only weakly to neutrinos. Hence the total scattering rate is
\begin{equation}
\Gamma_{\rm scattering} = \frac{\sum_\pi Q_\pi^2}{N_\pi} \sum_{f \,=\, e^-, \,e^+} \langle \sigma v(\pi_+ f \to \pi_+ f) n_f \rangle = \frac{1}{2} \frac{\varepsilon^2 \hat{e}^2 e^2  c_w^2 }{2\pi m_{A'}^4}\, (g_{e^-} + g_{e^+})\, \frac{45 \zeta (5)}{4\pi^2}\, T^5\,,
\end{equation}
where the ratio $\sum_\pi Q_\pi^2 / N_\pi = 1/2$ accounts for the fact that half of the $N_\pi = 8$ mesons have unit charge under the dark EM and the other half are neutral (this is true regardless of whether we take $Q$ or $Q'$ charges for the quarks). The relevant thermal average is $(2\pi)^{-3} \int d^3p_f p_f^2/ [\mathrm{exp}(p_f/T) + 1] = 45 \zeta(5)  T^5 / (4\pi^2)$, and $g_{e^-} = g_{e^+} = 2\,$. The hidden sector is therefore thermalized down to freeze-out as long as
\begin{equation}
\frac{5\zeta(5)}{4} \frac{T}{m_{\pi}} \Gamma_{\rm scattering} \gtrsim H
\end{equation}
at $T_{\rm fo}^{3\to2}$~\cite{Hochberg:2015vrg}. This condition can be rewritten as a lower bound on $\varepsilon$,
\begin{align} 
\varepsilon \gtrsim  \frac{16\pi^2 (x_{\rm fo}^{3\to2})^2}{15 \zeta(5) \hat{e} e c_w}& \Big( \frac{\sqrt{g_\ast}}{3\sqrt{10}}\Big)^{1/2}   \left( \frac{m_{A'}}{2m_\pi} \right)^2  \left( \frac{m_\pi}{M_{\rm Pl}} \right)^{1/2} \hspace{-2mm} \nonumber \\ \label{eq:eps_therm}
&= 9 \times 10^{-6}  \left( \frac{m_{A'}}{2m_\pi} \right)^2 \left( \frac{x_{\rm fo}^{3\to2}}{20} \right)^2 \left( \frac{\alpha}{\hat{\alpha}} \right)^{1/2}  \left( \frac{m_{\pi}}{200\;\mathrm{MeV}} \right)^{1/2},
\end{align}
where the mass splittings among the mesons have been neglected. We note that the elastic scattering of the DM triplet on hidden-charged mesons is extremely efficient at $T_{\rm fo}^{3\to2}$, i.e. \mbox{$\langle \sigma v\rangle_{LM\to LM} \,n_{\pi_+}^{\rm eq} \gg H$}. This remains true even if $\Delta_\pi$ is increased and $n_{\pi_+}^{\rm eq}\ll n_{\pi_0}$ at $x_{\rm fo}^{3\to2}$, maintaining kinetic equilibrium between the DM and the SM sector.

%%%%%%%%%%%%%%%%%%%%%%%%%%%%%%%%%%%%%%%%%%%%%%%%%%%%%%%%%%%%%%%%%%%%%%%%%%%%%%%
\subsection{Suppressed semi-annihilation to dark photon and annihilation to SM} \label{sec:semiann_ann}
In order for $3\to 2$ annihilations to drive DM freezeout, the semi-annihilation processes $\pi\pi \to \pi A'$ (mediated by the AAAV part of the WZW action that appears in the third line of Eq.~\eqref{eq:ChPT}) should be sufficiently suppressed. This cannot be achieved by arbitrarily decreasing $\hat{\alpha}$, as $\hat{\alpha}\hspace{0.25mm}\varepsilon^2$ is bounded from below by the thermalization requirement in Eq.~\eqref{eq:eps_therm}. Thus we restrict our analysis to $m_{A'} > 2m_\pi$, which, as discussed in Ref.~\cite{Hochberg:2015vrg}, ensures the kinematic suppression of semi-annihilations, including the effect of the thermal tail, as $Y_{\pi} Y_{A'} \lesssim Y_\pi^3$ with $Y\sim e^{- m/T}$ at freezeout.

In addition, we also require that dark meson annihilation to SM leptons is out of equilibrium at $T_{\rm fo}^{3\to2}$. The cross section for, e.g., $\pi_+ \pi_- \to f \bar{f}$ mediated by the exchange of neutral vectors ($\widetilde{V}_{A,B} = A', Z$) in the $s$-channel is
\begin{equation} \label{eq:SM_ann}
\sigma v(\pi_+ \pi_- \to f \bar{f}) = \frac{1}{6\pi } (s - 4m_\pi^2) \sum_{\widetilde{V}_A, \widetilde{V}_B} \frac{g_{\widetilde{V}_A \pi \pi } g^\ast_{\widetilde{V}_B\pi \pi } }{ (s - m^2_{\widetilde{V}_A})(s - m^2_{\widetilde{V}_B}) } (g_{\widetilde{V}_A f \bar{f}}^V \,g^{V \ast}_{\widetilde{V}_B f \bar{f}} + g_{\widetilde{V}_A f \bar{f}}^A \,g^{A \ast}_{\widetilde{V}_B f \bar{f}})\,,
\end{equation}
where we have assumed that $f$ is color neutral and neglected its mass. The exchange of $A'$ dominates, hence the thermally-averaged total annihilation rate is
\begin{equation}\label{eq:Gamma_ann}
\Gamma_{\rm annihilation} = \sum_{\ell \,=\, e,\, \mu} \frac{\sum_{\pi} Q_\pi^2}{N_\pi^2} \frac{m_\pi T}{\pi} \frac{\varepsilon^2 \hat{e}^2 e^2 c_w^2 Q_\ell^2}{(4m_\pi^2 - m_{A'}^2)^2}\, s\hspace{0.1mm} Y_{\rm DM}\,,
\end{equation}
where $s = 2 \pi^2 g_{\ast s} T^3/45$ and $Y_{\rm DM} \approx 4.1\times 10^{-10}\, (\mathrm{GeV} / m_{\pi})$. Notice that we have neglected the annihilation to SM charged pions. Performing the average $\sum_\ell \sum_\pi Q_\pi^2 / N_\pi^2 = 1/8$ and requiring $\Gamma_{\rm annihilation} \lesssim H$ at $T_{\rm fo}^{3\to2}$, we obtain the region above the dotted gray curve in the upper left portion of the $(m_{A'}, \varepsilon)$ plane in Fig.~\ref{fig:summary}. Here we have ignored the possible mass splittings among the mesons, which would make the charged mesons more Boltzmann suppressed at $x_{\rm fo}^{3\to 2} \approx 20$, mildly shifting the curve upwards.

Finally, we comment that the decays of $A'$ do not affect any of the discussions in this section, as $A'$ freezes out with a very suppressed abundance and has a short lifetime. The $A'$ population decays dominantly to hidden charged mesons. Even in the narrow sliver of parameter space $2 m_{\pi_0} < m_{A'} < 2 m_{\pi_+}$ where this channel is forbidden by kinematics, $A^\prime$ decays to SM particles with $\Gamma_{A'} \sim \alpha c_w^2 \varepsilon^2 m_{A'}$, corresponding to a lifetime $\tau_{A'} \sim 10^{-9}\;\mathrm{s}\; (10^{-6}/\varepsilon)^2\hspace{0.2mm} (0.1\;\mathrm{GeV}/m_{A'})$, which is orders of magnitude too rapid to affect Big Bang nucleosynthesis (BBN). 

%%%%%%%%%%%%%%%%%%%%%%%%%%%%%%%%%%%%%%%%%%%%%%%%%%%%%%%%%%%%%%%%%%%%%%%%%%%%%%%%%%%%%%%
%%%%%%%%%%%%%%%%%%%%%%%%%%%%%%%%%%%%%%%%%%%%%%%%%%%%%%%%%%%%%%%%%%%%%%%%%%%%%%%%%%%%%%%%
%%%%%%%%%%%%%%%%%%%%%%%%%%%%%%%%%%%%%%%%%%%%%%%%%%%%%%%%%%%%%%%%%%%%%%%%%%%%%%%%%%%%%%%%
\section{Constraints and signatures} 
\label{sec:constraints}
Our framework admits a wide variety of signals on several fronts, spanning early Universe cosmology, DM indirect detection, DM self-interactions, and dark photon searches. We now discuss these constraints and signatures in turn before summarizing them in Fig.~\ref{fig:summary}.

\subsection{$\eta$ decays}\label{sec:eta_decays}
As discussed in Section \ref{subsec:etadecay}, the $\eta$ decays even when the isospin symmetry is exact. If $m_\eta < 2m_\mu$ the dominant decay channel is $\eta \to 4e$, whereas $\eta\to \mu\mu$ dominates for heavier masses. The lifetime spans a large range of values in the allowed parameter space, but is generally much longer than one second, i.e. $\eta$ decays after BBN, and shorter than the age of the Universe, so that $\eta$ cannot be a significant component of the present DM.  

\begin{figure}[t!]
\begin{center}
{\includegraphics[width=0.5\textwidth]{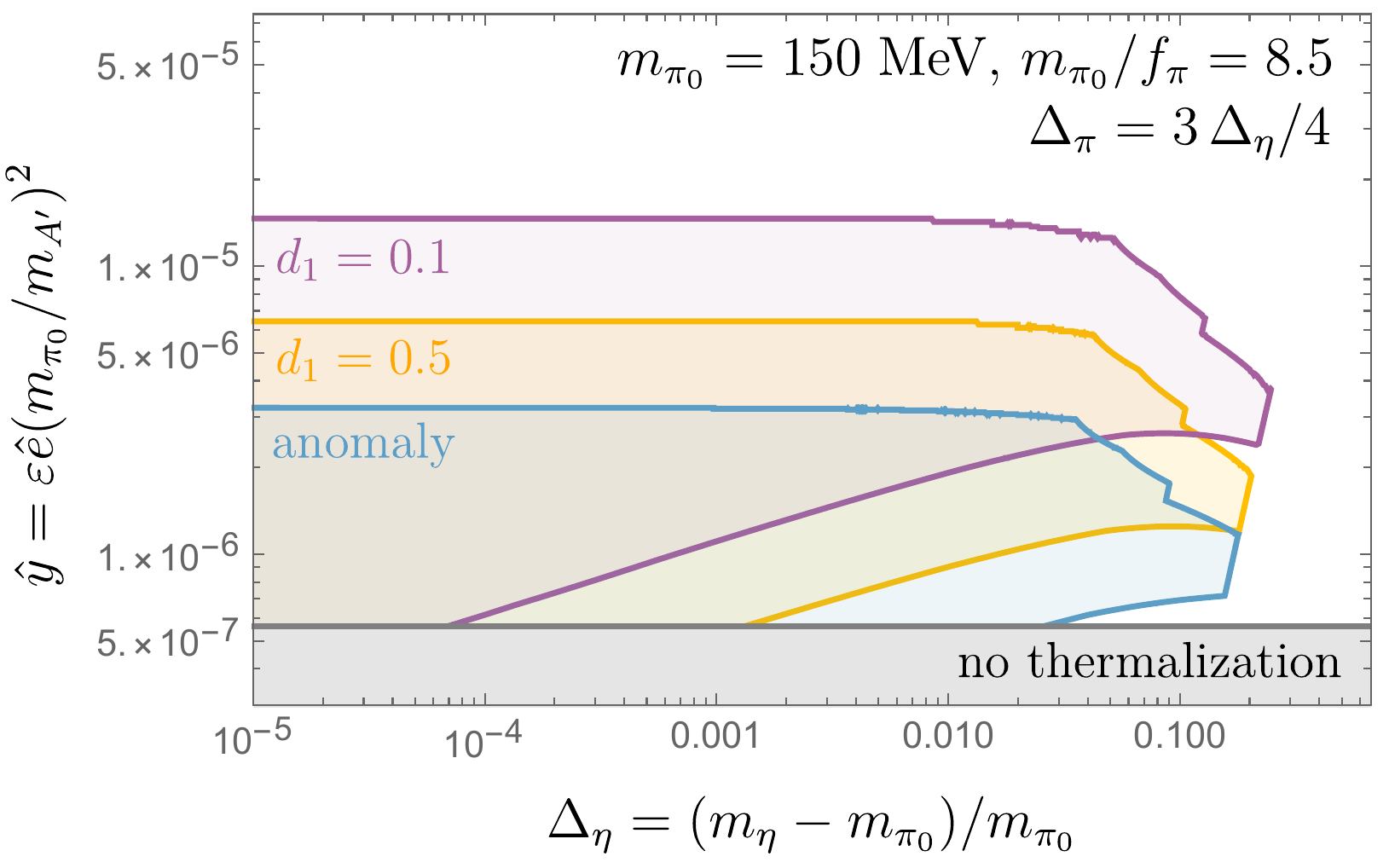}} 
\caption{\label{fig:etadecay} Exclusion regions from $\eta$ decay. We have chosen parameters so that only the $\eta\to 4e$ decay channel is relevant in setting the bounds. In this case both the $\eta$ lifetime and the kinetic decoupling depend on the combination $\hat{y} = \varepsilon \hat{e} (m_{\pi_0} /m_{A'})^2$, shown on the vertical axis. The variable $\Delta_\eta$ on the horizontal axis mainly controls the freezeout abundance of $\eta$, under the assumption of leading-order masses for the mesons.} 
\end{center}
\end{figure}

We derive constraints on the decay of $\eta$ in the early Universe based on the results presented in Ref.~\cite{Poulin:2016anj} (see also Ref.~\cite{Slatyer:2016qyl}), which provided limits on the energy fraction of a decaying particle as a function of its lifetime. In Fig.~\ref{fig:etadecay} we present bounds assuming $m_{\pi_0} = 150$~MeV and $m_{\pi_0}/f_\pi = 8.5$, as in Fig.~\ref{fig:Boltzmann_2}, for which only the $\eta\to 4e$ decay channel is relevant. We consider both anomalous and non-anomalous decay, with two different choices of the $d_1$ coefficient in Eq.~\eqref{eq:eta_decay_op} for the latter case. The exclusions are shown as regions in the $(\Delta_\eta, \hat{y})$ plane, whose shapes reflect Fig.~5 of Ref.~\cite{Poulin:2016anj}. The parameter $\Delta_\eta$ controls the mass of $\eta$ and therefore its freezeout abundance before it decays (see Fig.~\ref{fig:Boltzmann_2}), as well as its lifetime to a moderate extent. The combination $\hat{y} \equiv \varepsilon \hat{e} (m_{\pi_0} /m_{A'})^2$ determines both the $\eta$ lifetime and the kinetic decoupling; the latter has a mild impact on the $\eta$ 
freezeout abundance.\footnote{Our $\hat{y}$ is related to the commonly-used 
variable $y$~\cite{Izaguirre:2015yja} by $\hat{y}^2 =  4\pi y$.} 

For $\tau \lesssim 10^{12}$~s, constraints come from BBN observables and CMB spectral distortions, whereas CMB anisotropies provide the leading sensitivity for longer lifetimes. The bounds are strongest when the lifetime matches the timescale of recombination, $\tau\sim 10^{13}$~s, corresponding for example to 
$\hat{y}\sim 10^{-6}$ for decay mediated by the anomaly. Larger values of $\Delta_\eta$ lead to larger $\eta\,$--$\,\pi_0$ mass splitting and therefore to a greater suppression of the $\eta$ freezeout density, relaxing such constraints. As seen in Fig.~\ref{fig:etadecay}, the bounds completely disappear for $\Delta_\eta \gtrsim 0.2$, corresponding to $n_\eta / n_{\pi_0}\lesssim 10^{-11}$, for which the lifetime becomes unconstrained~\cite{Poulin:2016anj,Slatyer:2016qyl}. For $m_\eta > 2 m_\mu$ the exclusion from $\eta$ decays has a different parametric dependence (see the middle panel of Fig.~\ref{fig:summary} and Section~\ref{sec:summary}).

%%%%%%%%%%%%%%%%%%%%%%%%%%%%%%%%%%%%%%%%%%%%%%%%%%%%%%%%%%%%%%%%%%%%%%%
\subsection{Dark matter decays} 
\label{sec:DM_decays}
Next, we discuss constraints and prospects for DM decay. We mostly focus on the region $m_{\pi_0} < 2 m_\mu$, where $\pi_0$ dominantly decays to $4e$. As we learned in the previous sections, DM is composed of the $K_0$ and $\overline{K}_0$, which are stable, and the $\pi_0$, which decays through its mixing with $\eta$, proportional to the small isospin-breaking parameter $\delta$. For simplicity, in this subsection we quote lifetime bounds and projected sensitivities for a decaying species that constitutes all of DM. Our numerical results take into account that $\pi_0$ only forms a fraction $r_{\pi_0}$ of the DM density, and as a consequence the limits are weaker by a factor $1/r_{\pi_0}\hspace{0.15mm}$. As we saw in Section~\ref{sec:pi0_K0_splitting}, the $\pi_0\,$--$\,K_0$ mass splitting can cause a depletion of the neutral kaons before the time of recombination. For realistic values of $\delta \lesssim 10^{-5}$, however, the effect is mild, and $r_{\pi_0}\gtrsim 1/3$ is appropriate for evaluating CMB constraints. Later, after structures form up-scattering equilibrates the three DM subcomponents, so that $r_{\pi_0} = 1/3$ at the present epoch. 

The leading CMB sensitivity on DM decays comes from the anisotropies of the 
angular power spectra~\cite{Slatyer:2016qyl,Poulin:2016anj}. For $O(100)$ MeV DM decaying to $e^+ e^-$, the bound is $\tau_{\,\rm DM} \gtrsim 2 \times 10^{25}$~s~\cite{Slatyer:2016qyl}, which we adopt here. A precise constraint would need to take into account the difference between our four-body decay and the two-body $e^+ e^-$ injection assumed 
in Refs.~\cite{Poulin:2016anj,Slatyer:2016qyl}, likely resulting in an $O(1)$ correction. Recently, improved bounds on DM decay were obtained from the measurement of the intergalactic medium temperature derived from the Lyman-$\alpha$ forest~\cite{Liu:2020wqz}. Under conservative assumptions, this bound is mildly weaker than the CMB one~\cite{Slatyer:2016qyl} for $O(100)$ MeV DM.

There also exist indirect detection bounds on DM decay in the present epoch. The $O(100)$ MeV masses relevant to our setup fall in the so-called ``MeV gap'' of gamma astronomy, where the current constraints from gamma ray measurements are relatively weak. The leading bounds come from the COMPTEL experiment. Reference~\cite{Essig:2013goa} used COMPTEL data to derive very conservative limits, without any background subtraction, 
of $\tau_{\,\rm DM} \gtrsim 10^{25}$~s for DM decay into $e^+ e^-$ with the emission of final state radiation.\footnote{Note that if DM in this mass range decays dominantly to SM final states containing photons or neutral pions, such as $\gamma\gamma$, $\gamma \pi_0$, or $\pi_0 \pi_0$, then the gamma ray constraints are stronger~\cite{Boddy:2015efa}. However, the dark quarks do not couple to the visible photon, whereas the decay of the $\pi_0$ to two SM pions, either neutral or charged, violates $CP$ symmetry.} These constraints are affected by uncertainties on the galactic DM density profile. However, since the signal rate from DM decay scales as $\rho_{\rm DM}$, compared to $\rho_{\rm DM}^2$ for annihilation, uncertainties on the DM profile have smaller impact on indirect detection limits for decaying DM compared to annihilating DM. Recently, new constraints on annihilation of sub-GeV DM were derived from INTEGRAL data, by exploiting the lower-energy X-rays produced via inverse Compton scattering~\cite{Cirelli:2020bpc}. A rough extrapolation to the case of decaying $O(100)$~MeV DM indicates a sensitivity comparable to that of COMPTEL \cite{Essig:2013goa}. Stronger bounds, $\tau_{\,\rm DM} \gtrsim \mathrm{few} \times 10^{26}$~s, were derived in Ref.~\cite{Boudaud:2016mos} from the measurement of electrons and positrons in 
the interstellar medium by Voyager~1. These $e^\pm$ constraints come with the usual caveats about uncertainties in the modeling of cosmic ray propagation in the galaxy.

In the future, improved gamma ray limits are expected from the proposed AMEGO observatory~\cite{McEnery:2019tcm}. To estimate the AMEGO sensitivity to $\pi_0$ decays we use the results of Ref.~\cite{Bartels:2017dpb}, which performed a detailed study of the future gamma ray reach on annihilating DM in the MeV--GeV range, using eASTROGAM~\cite{DeAngelis:2016slk} as benchmark. Taking into account that for $m_{\rm DM} = O(100)$~MeV the ultimate AMEGO reach will be approximately a factor~2 weaker than for eASTROGAM~\cite{McEnery:2019tcm,DeAngelis:2016slk}, we apply the improvement factor for annihilating DM derived in Ref.~\cite{Bartels:2017dpb} to our decaying DM, and estimate that the AMEGO sensitivity will be $\sim 50$ times stronger than the current COMPTEL limit from Ref.~\cite{Essig:2013goa}, reaching 
$\tau_{\,\rm DM} \gtrsim 5 \times 10^{26}$~s. Thus AMEGO will become competitive with, and even surpass, the limits from Voyager~1 reported in Ref.~\cite{Boudaud:2016mos}, providing a strongly complementary exploration of the parameter space.

We emphasize that the above indirect detection estimates are based on existing results for DM decaying (and annihilating) to $e^+ e^-$, whereas the most relevant decay channel here is $4e$, for which dedicated studies are not available. While a detailed treatment of this channel would be interesting, we expect that it would only induce $O(1)$ corrections to our estimates. For example, Ref.~\cite{Bartels:2017dpb} found the gamma ray constraint on a related process, 
$\mathrm{DM}\;\mathrm{DM}\to \phi\,\phi,\, \phi\to e^+e^-$, to be about $4$ times weaker than for $\mathrm{DM}\;\mathrm{DM}\to e^+e^-$; the limit on our direct four-body process would fall somewhere in between. 

When $\pi_0$ is heavier than the dimuon threshold, as in the bottom panel of Fig.~\ref{fig:summary}, for indirect detection bounds we make use of the results for the $\mu\mu$ channel presented in 
Refs.~\cite{Boudaud:2016mos,Bartels:2017dpb}. For the CMB constraints, as suggested in Ref.~\cite{Poulin:2016anj}, we assume that, since neutrinos carry away an average fraction $x \approx 2/3$ of the energy, the bound on 
$\tau_{\,\mathrm{DM} \to \mu\mu}$ is roughly given by $(1-x)$ times the bound on $\tau_{\,\mathrm{DM} \to ee}$~\cite{Slatyer:2016qyl} evaluated 
at $(1-x)m_{\mathrm{DM}}$. 

%%%%%%%%%%%%%%%%%%%%%%%%%%%%%%%%%%%%%%%%%%%%%%%%%%%%%%%%%%%%%%%%%%%%%%%%%%%%%%%%
\subsection{Dark matter self-interactions}
One of the salient features of the SIMP paradigm is a large DM self-interaction 
cross section. The leading contributions to $\pi \pi \to \pi \pi$ scattering arise 
from the kinetic and the mass terms in the chiral Lagrangian, 
\begin{equation} \label{eq:self_int}
- \frac{r_{abcd}}{24 f_\pi^2} \pi^a \pi^b \partial_\mu \pi^c \partial^\mu \pi^d  + \frac{\overline{m}_\pi^2 c_{abcd}}{48 f_\pi^2} \pi^a \pi^b \pi^c \pi^d  + \frac{\overline{m}_\pi^2 \Delta m}{48 f_\pi^2 m} \pi^a \pi^b \pi^c \pi^d\, \mathrm{Tr}[\mathrm{diag}(1,0,0) \lambda^a \lambda^b \lambda^c \lambda^d], 
\end{equation}
where 
$r_{abcd} = 2 ( f^{ace} f^{bde} + f^{bce} f^{ade})$ and $c_{abcd} = 
\frac{4}{3N_f} (\delta^{ab} \delta^{cd} + \delta^{ac} \delta^{bd} + 
\delta^{ad}\delta^{bc}) + \frac{2}{3} (d^{abe} d^{cde} + d^{ace}d^{bde} +
d^{ade}d^{cbe})$\,, with $f^{abc}$ the $SU(N_f)$ structure constants and $d^{abc}$ its fully symmetric symbols.\footnote{For $N_f = 2,3$ one can obtain the simpler form $c_{abcd} = \frac{2}{3} (\delta^{ab} \delta^{cd} + \delta^{ac} \delta^{bd} + \delta^{ad}\delta^{bc})$.} We also defined $\overline{m}_\pi^2 = 2 B m$ and ignored the small effect of isospin breaking.

As discussed in Section~\ref{sec:pi0_K0_splitting}, the very small mass splitting between $\pi_0$ and neutral kaon states induced by isospin breaking can cause an $O(1)$ depletion of the latter population, but subsequent up-scattering of $\pi_0$ in halos re-equilibrates the densities of the three DM subcomponents. Therefore we assume the current DM abundance to be an equal admixture of $\pi_0, K_0,$ and $\overline{K}_0$, for which the self-scattering cross section is~\cite{Hochberg:2014kqa}
\begin{equation} \label{eq:sigma_ave}
\sigma_{\rm average} = \frac{m_{\pi}^2}{256 \pi f_{\pi}^4} \frac{c^2 +  r^2 / 9}{2 N_\pi^2}\,,
\end{equation}
where, defining $R_{abcd} \equiv r_{abcd} + r_{abdc} + r_{bacd} + r_{badc}\,$, we find
\begin{equation}
r^2 = \sum R_{abcd}^2 = 192 N_f^2 (N_f^2 -1)\,, \qquad c^2 = \sum c_{abcd}^2 = \frac{8}{3N_f^2} (N_f^6 - 7 N_f^4 + 24 N_f^2 - 18)\,,
\end{equation}
and $N_\pi = N_f^2-1$.\footnote{The final term in Eq.~\eqref{eq:self_int} does not contribute to triplet self-scattering. Note also that $\sum c_{abcd} R_{abcd} = 0$.} For our DM triplet, the cross section is given by Eq.~\eqref{eq:sigma_ave} with $N_f = 2$. The rate of self-scatterings is proportional to $\sigma_{\rm average}/m_{\rm DM}$, for which we obtain
\begin{equation} \label{eq:self_scatt_degen}
\frac{\sigma_{\rm average}}{m_{\pi_0}} = \frac{23 \hspace{0.2mm}m_{\pi_0}}{384 \pi f_{\pi}^4} \approx 4.2\;\frac{\mathrm{cm}^2}{\mathrm{g}} \left( \frac{200\;\mathrm{MeV}}{m_{\pi_0}} \right)^3 \left( \frac{m_{\pi_0}/f_\pi}{9.5} \right)^4\,.
\end{equation}
For the three benchmark scenarios presented in Fig.~\ref{fig:summary}, this quantity ranges from $\sim 2$ to $\sim 6$ cm$^{2}/\mathrm{g}$, exhibiting some tension with the $O(1)$ cm$^2$/g constraints from the Bullet cluster~\cite{Randall:2007ph} and halo shapes~\cite{Rocha:2012jg,Peter:2012jh}. In view of the ongoing debate concerning the CDM small-scale puzzles and their possible DM explanations (see Ref.~\cite{Tulin:2017ara} for a review), we leave a conclusive statement about the viability of this cross section to the future. Incidentally, we note that the cross section in Eq.~\eqref{eq:self_scatt_degen} also approximately applies to the setup of Ref.~\cite{Hochberg:2018vdo}, where the DM is an $SU(2)$ triplet in a $4$-flavor hidden QCD theory.

If the neutral kaon and pion were split by an amount $\Delta_K \gtrsim v_0^2$, where $v_0 \sim 0.003$ for the DM velocity on cluster scales, then $K_0 \overline{K}_0 \to \pi_0 \pi_0$ scattering would very efficiently deplete the kaons in the early Universe, and the current DM population would be dominated by $\pi_0$'s. These only undergo elastic $\pi_0 \pi_0 \to \pi_0 \pi_0$ scattering, as up-scattering to heavier mesons is kinematically forbidden. Since $r_{aaaa} = 0$, the cross section would be reduced to
\begin{equation} \label{eq:self_scatt_pi0}
\frac{\sigma}{m_{\pi_0}} = \frac{m_{\pi_0}}{128 \pi f_{\pi}^4} \approx 0.55 \;\frac{\mathrm{cm}^2}{\mathrm{g}} \left( \frac{200\;\mathrm{MeV}}{m_{\pi_0}} \right)^3 \left( \frac{m_{\pi_0}/f_\pi}{9.5} \right)^4\,,
\end{equation}
easily in agreement with constraints for our parameters. This observation was already made in Ref.~\cite{Hochberg:2014kqa}, and the cross section in Eq.~\eqref{eq:self_scatt_pi0} was assumed in Ref.~\cite{Berlin:2018tvf}. However, from Eq.~\eqref{eq:DeltaK} we find that the above scenario requires $\delta \gtrsim 10^{-2}$, corresponding to $\pi_0$ lifetimes that are orders of magnitude shorter than the experimental bounds. Thus, our results illustrate quantitatively that it is nontrivial to realize a setup where a single real meson constitutes all of SIMP DM and has a sufficiently long lifetime, motivating further investigation into this question.

Finally, we remind the reader that the DM self-scattering cross section decreases as the number of hidden colors is increased~\cite{Hochberg:2014kqa}: for fixed $m_{\pi}, N_f$ and larger $N_c$ the correct relic density is obtained at larger $f_\pi$, leading to a suppression $\sigma \propto N_c^{\,-4/5}$. Furthermore, a larger $f_\pi$ allows for heavier mesons within the perturbativity bound $m_\pi < 4 \pi f_\pi$, which affords a further reduction of the self-scattering cross section. However, the large$\,$-$N_c$ scaling of the vector meson masses, $m_V \sim 4\pi f_\pi / N_c^{1/2}\,$, indicates that when $N_c$ is increased the effects of these resonances become increasingly important~\cite{Berlin:2018tvf}, making pure chiral perturbation theory inapplicable. In addition, heavier mesons generically have shorter lifetimes, requiring a higher degree of symmetry to avoid conflict with experimental constraints.

%%%%%%%%%%%%%%%%%%%%%%%%%%%%%%%%%%%%%%%%%%%%%%%%%%%%%%%%%%%%%%%%%%%%%%%%%%%%%%%%%%%%
\subsection{Dark photon searches at laboratory experiments}
We now discuss existing and future constraints on the $A'$, which couples dominantly to dark quarks and therefore decays dominantly to invisible final states. These bounds and projections are collected in Fig.\,\ref{fig:summary}.

The strongest constraint comes from the BaBar search for $e^+ e^- \to \gamma A'$, $A'\to$~invisible, based on $53$ fb$^{-1}$ of data~\cite{Lees:2017lec} (see Refs.~\cite{Izaguirre:2013uxa,Essig:2013vha} for pioneering reinterpretations of preliminary BaBar results). The energy of the photon in the center-of-mass frame is given by $E_\gamma^\ast = (s - m_{A'}^2)/(2\sqrt{s}\,)$, where $\sqrt{s}$ corresponds to the $\Upsilon(2S)$ and $\Upsilon(3S)$ resonances, hence the experimental requirement $E_\gamma^\ast \gtrsim 1.8\;\mathrm{GeV}$ implies sensitivity to $m_{A'}\lesssim 8\;\mathrm{GeV}$. In addition, we show the reach attainable in the near future by Belle II with $20$ fb$^{-1}$ of data~\cite{Kou:2018nap}. At low-energy lepton colliders, performing dark spectroscopy is also possible~\cite{Hochberg:2015vrg,Hochberg:2017khi}.

For small $m_{A'} \lesssim \mathrm{GeV}$, fixed-target experiments provide additional sensitivity. NA64 has performed a search for invisibly-decaying dark photons, exploiting bremsstrahlung production from a $100$~GeV electron beam that scatters in an active dump and searching for the missing energy signature~\cite{NA64:2019imj}. This result is based on $2.84\times 10^{11}$ electrons on target, expected to increase up to $5\times 10^{12}$ in the future~\cite{Beacham:2019nyx}. Looking further ahead, the proposed LDMX experiment~\cite{Akesson:2018vlm} will be able to extend the reach for light invisible dark photons by searching for the missing momentum signature. We consider two different projections for the long-term LDMX reach: an extended run beyond Phase I as discussed in Ref.~\cite{Akesson:2018vlm} (see Fig.~79 therein), and the ``ultimate'' sensitivity attainable with a $16$~GeV electron beam at the proposed eSPS facility at CERN~\cite{Beacham:2019nyx}. For completeness, we also mention that the proposed KLEVER experiment could provide interesting sensitivity to lighter dark photons in the $m_{A'}\sim 0.1\hspace{0.2mm}$--$\,0.3$~GeV window~\cite{Beacham:2019nyx} by searching for $K_L \to \pi_0 A'$ decays.

For $m_{A'} \gtrsim 8$~GeV the leading constraint comes from electroweak precision tests (EWPT), where the dark photon induces a 
shift in the $Z$ mass and corrections to its couplings to SM fermions. We show these constraints as reported in Ref.~\cite{Curtin:2014cca}. In combination with the thermalization bound, Eq.~\eqref{eq:eps_therm}, EWPT rule out the region $m_{A'} \gtrsim 50$~GeV. For $m_{A'}\ll m_Z$ the main effect is 
the correction to the $Z$ mass, 
$m^2_Z \simeq m_{Z_0}^2 ( 1 + s_w^2 \varepsilon^2)$~\cite{Hook:2010tw}. 
We also show the expected improvement after the completion of the LHC program, assuming in particular an $8\,(440)$~MeV precision 
on $m_W\,(m_t)$ and a reduction of the uncertainty on $\Delta\alpha_{\rm had}^{(5)}(m_Z)$ by a factor $2$, with the refined measurement of $m_W$ being largely responsible for the increased sensitivity of the fit~\cite{Curtin:2014cca}. We also consider recently derived, competitive constraints from deep inelastic scattering (DIS) at HERA~\cite{Kribs:2020vyk}, which like EWPT are insensitive to the decay pattern of the dark photon.

In addition, we display the bound from a monophoton search at DELPHI, originally analyzed as a constraint on DM coupling to electrons in Ref.~\cite{Fox:2011fx} and later recast to the dark photon case~\cite{Ilten:2018crw}. Monojet searches at hadron colliders also probe the invisibly-decaying $A'$. In the region $m_{A'}\lesssim m_Z$ the strongest bounds are still those derived from CDF data in Ref.~\cite{Shoemaker:2011vi}. ATLAS~\cite{Aaboud:2017phn} and CMS~\cite{Sirunyan:2017jix} searches (based on $36$ fb$^{-1}$ of $13$~TeV data) give comparable but weaker constraints due to much stricter selection criteria enforced in particular by trigger requirements, which imply a loss of sensitivity to the soft signal considered here~\cite{Shoemaker:2011vi}. In the light of this, it appears challenging for the LHC monojet searches to improve on EWPT limits, even in the high-luminosity phase. In the CMS monophoton search~\cite{Sirunyan:2018dsf} the transverse momentum requirements are only moderately softer compared to the monojet channel, so we do not expect competitive sensitivity in our scenario. We remark that when the $A'$ decays dominantly to the dark sector, i.e. for $\hat{\alpha} \gg \alpha\, \varepsilon^2$, the constraints from invisible final states are insensitive to $\hat{\alpha}$.

Finally, searches in the $\ell^+\ell^-$ final states (with $\ell =e$ or $\mu$) at lepton and hadron colliders~\cite{Lees:2014xha,Aaij:2019bvg,Sirunyan:2019wqq} rival EWPT for the best current sensitivity in the $m_{A'} \gtrsim 8$~GeV region, despite the strongly-suppressed branching ratio to SM particles of our dark photon. BaBar searched for $e^+ e^- \to \gamma A'$, $A^\prime \to ee, \mu\mu$ in the mass range $0.02\;\mathrm{GeV} <  m_{A'} < 10.2~\mathrm{GeV}$, using 514 fb$^{-1}$ of data~\cite{Lees:2014xha}, whereas LHCb has recently searched for $q\bar{q}\to A' \to \mu\mu$ in the range $2m_{\mu} <  m_{A'} < 70~\mathrm{GeV}$~\cite{Aaij:2019bvg}. In our scenario these analyses are only competitive for $m_{A'} \gtrsim 8$~GeV, as the BaBar monophoton constraint is far stronger for smaller masses. In addition, CMS has recently performed a search for the dimuon signal in the $11.5\;\mathrm{GeV} <  m_{A'} < 200~\mathrm{GeV}$ region; for $m_{A'} < 45$~GeV, as relevant here, the search employs data scouting to enhance its sensitivity~\cite{Sirunyan:2019wqq}.

%%%%%%%%%%%%%%%%%%%%%%%%%%%%%%%%%%%%%%%%%%%%%%%%%%%%%%%%%%%%%%%%%%%%%%%%%%%%%
\subsection{Dark matter direct detection and annihilation to SM}
In our setup the $(\pi_0, K_0, \overline{K}_0)$ 
triplet is neutral under dark EM, 
whereas the charged doublet states $(\pi_\pm, K_\pm)$ have exponentially 
suppressed relic abundances, as required to avoid cosmological constraints 
from $\eta$ decay. As a consequence, the signal of DM scattering on electrons 
via dark photon exchange~\cite{Hochberg:2015vrg} is beyond the foreseeable experimental reach. 
 
For the same reasons, DM annihilation to SM leptons is extremely suppressed, avoiding any constraints from the CMB and indirect detection. Note that the annihilation cross section, Eq.~\eqref{eq:SM_ann}, is $p$-wave suppressed, implying that such constraints are very weak even for degenerate mesons. 

%%%%%%%%%%%%%%%%%%%%%%%%%%%%%%%%%%%%%%%%%%%%%%%%%%%%%%%%%%%%%%%%%%%%%%%%%
\subsection{Parameter space overview} 
\label{sec:summary}
\begin{figure}[t]
\begin{center}
\includegraphics[width=0.68\textwidth]{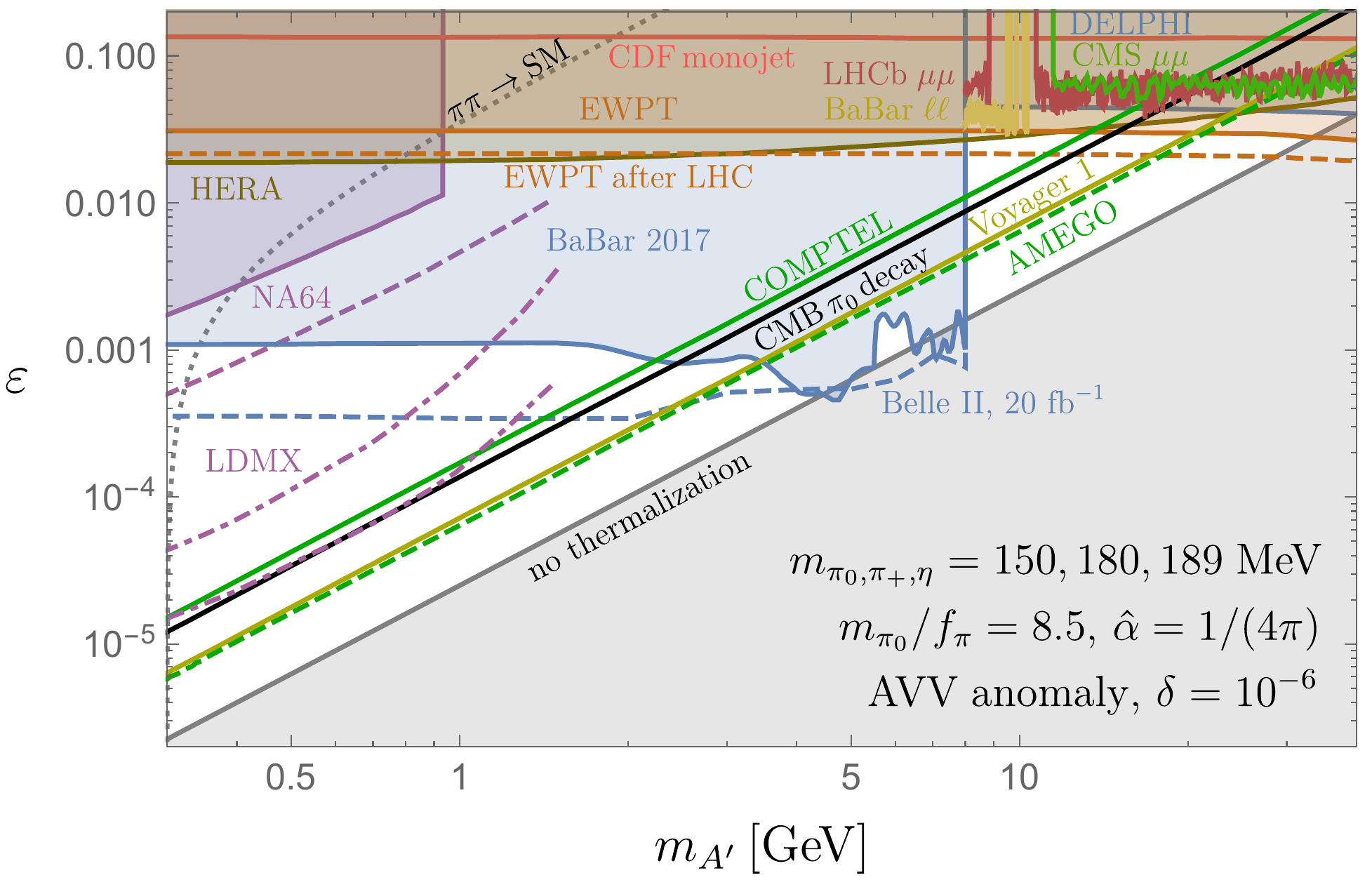}

\vspace{2mm}
\includegraphics[width=0.68\textwidth]{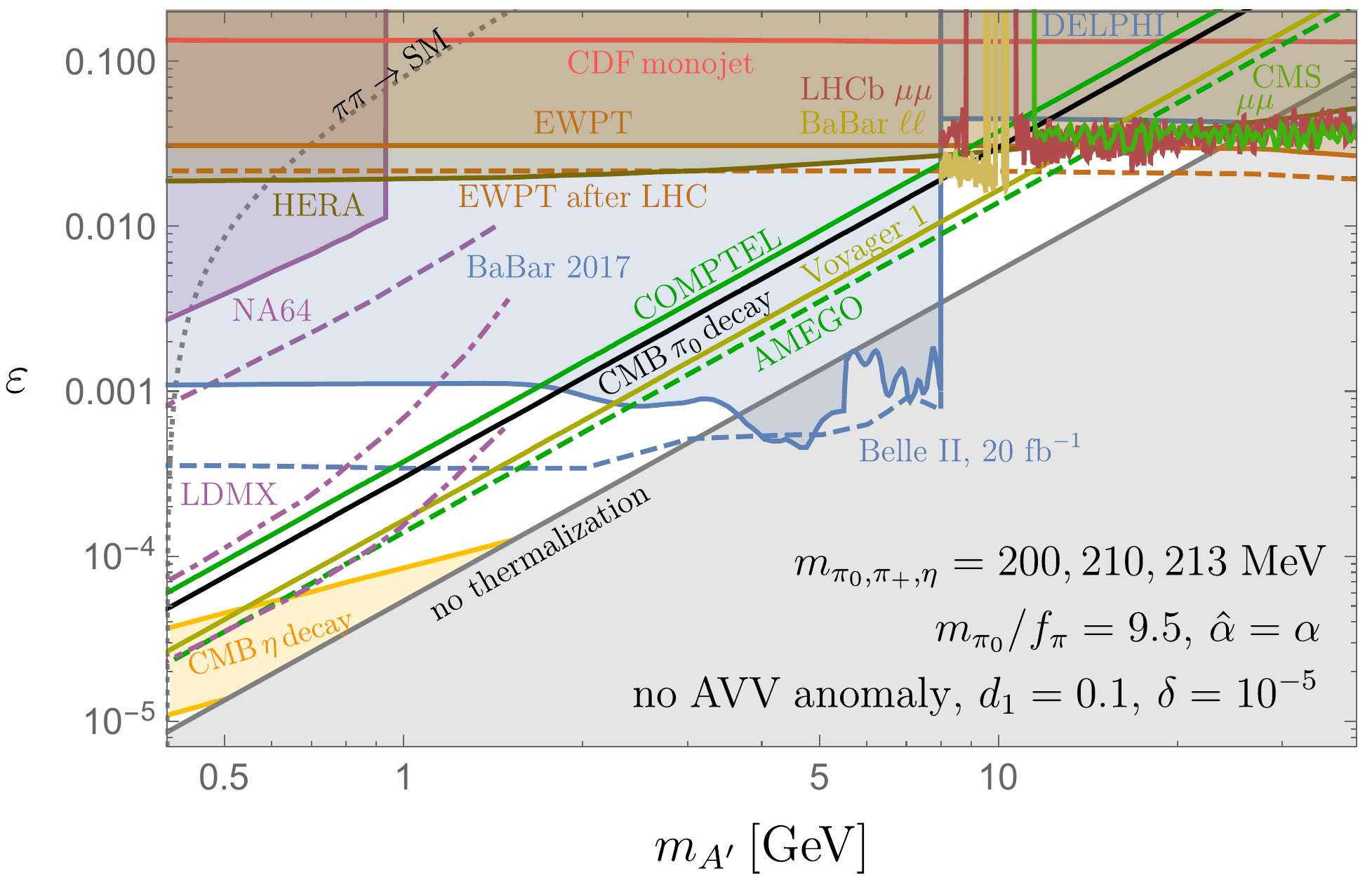}

\vspace{2mm}
\includegraphics[width=0.68\textwidth]{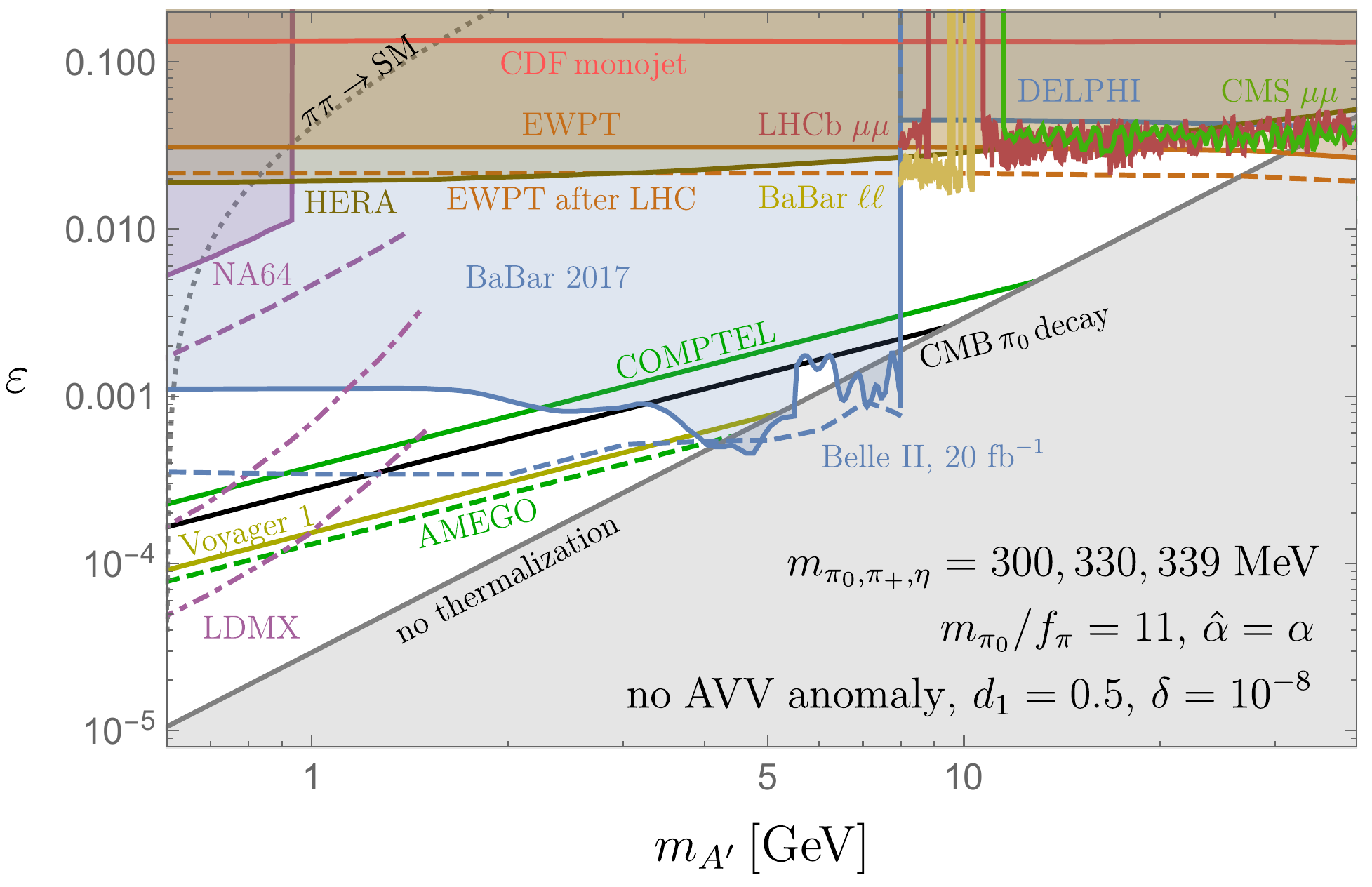}
\caption{\label{fig:summary} Summary of existing and projected constraints. See the text in Section~\ref{sec:summary} for details.} 
\end{center}
\end{figure}

We present in Fig.~\ref{fig:summary} the current and projected constraints in the $(m_{A'}, \varepsilon)$ parameter space based on the analysis in the previous subsections. Each of the three panels in this figure corresponds to a different representative benchmark for the meson masses and couplings.

{\it Top panel:} We consider the benchmark corresponding to Fig.~\ref{fig:Boltzmann_2}, with $Q$ charges for hidden EM (i.e., the AVV anomaly is present), $\hat{\alpha} = 1/(4\pi)$, and isospin-breaking parameter $\delta = 10^{-6}$. The region shaded in gray is not viable for SIMP DM, as the thermalization condition Eq.~\eqref{eq:eps_therm} is not met if $\hat{y} = \varepsilon \hat{e} (m_{\pi_0}/m_{A'})^2$ is too small. In the upper left region, above the dotted gray curve labeled $\pi\pi \to \mathrm{SM}$, direct annihilation to SM can be active at $3\to 2$ freezeout, as discussed after Eq.~\eqref{eq:Gamma_ann}, signaling a transition to a different regime. We recall that this curve was computed assuming degenerate mesons and would shift mildly upwards if mass splittings are included. Since $m_\eta < 2 m_\mu$ the $\eta$ decays to $4e$, but the chosen mass splitting parameter $\Delta_\pi = 0.2$ is large enough that the signal is below the Planck sensitivity on CMB anisotropies, see Fig.~\ref{fig:etadecay}. The constraints on DM decays, i.e. $\pi_0 \to 4e$ with lifetime given by Eq.~\eqref{eq:pi0_4e_anom}, are depicted as lines of constant $\hat{y}$, corresponding to $\varepsilon \propto m_{A'}^2$. For laboratory tests of the dark photon, we show regions excluded by monophoton searches at BaBar~\cite{Lees:2017lec} and DELPHI~\cite{Ilten:2018crw} (shaded in blue), as well as the projected near-term reach of Belle II~\cite{Kou:2018nap} (dashed blue curve). We also display the EWPT bound (orange-shaded region) together with the improvement expected by the end of HL-LHC~\cite{Curtin:2014cca} (dashed orange), and the exclusion from DIS at HERA~\cite{Kribs:2020vyk} (brown-shaded region). The best current monojet constraint is based on CDF data~\cite{Shoemaker:2011vi} (red), whereas CMS~\cite{Sirunyan:2017jix} rules out $\varepsilon \gtrsim 0.3$, which lies outside the range of the figure. For fixed-target experiments, we show the region already excluded by NA64~\cite{NA64:2019imj} (shaded in purple), the projected reach with $5\times 10^{12}$ electrons on target~\cite{Beacham:2019nyx} (dashed purple), and the projected sensitivity of LDMX in two possible scenarios (dot-dashed purple curves), namely an extended run beyond Phase I~\cite{Akesson:2018vlm} and a version with $16\;\mathrm{GeV}$ electron beam~\cite{Beacham:2019nyx}. Notice that while the collider bounds depend weakly on $m_{A'}$ (within the kinematically accessible range), the bremsstrahlung $A'$ production relevant to fixed-target experiments has a strong dependence on $m_{A'}$, resulting in the rather steep shape of the NA64 and LDMX curves. Finally, we show the current constraints derived from $A'\to \ell\ell$ at BaBar~\cite{Lees:2014xha} (dark yellow), as well as from $A'\to \mu\mu$ at LHCb~\cite{Aaij:2019bvg} (dark red) and CMS~\cite{Sirunyan:2019wqq} (light green). Although $\sigma \times \mathrm{BR} \propto \varepsilon^4$, as opposed to $\varepsilon^2$ for invisible final states, the strong sensitivity of the dilepton searches compensates for the suppression.
For the sake of readability we do not show the BaBar and LHCb limits for $m_{A'} < 8$~GeV, where they are much weaker than the BaBar monophoton bound.

{\it Middle panel:} We adopt the benchmark corresponding to the left panel of Fig.~\ref{fig:Boltzmann}, with the additional assumption of $Q'$ hidden electric charges so that the AVV anomaly is absent. The decays of $\eta$ and $\pi_0$ are then mediated by the operator in Eq.~\eqref{eq:eta_decay_op}, where we set $d_1 = 0.1$ and $\hat{\alpha} =\alpha$. The isospin-breaking parameter is fixed to $\delta = 10^{-5}$. In contrast with the top panel, here the meson mass splittings are smaller ($\Delta_\pi = 0.05$), and the $\eta$ density is not sufficiently suppressed to completely avoid CMB constraints (shaded in yellow). Since the $\eta$ mass is (just) above the $2m_\mu$ threshold, the CMB exclusion is approximately delimited by $\varepsilon \propto m_{A'}$ lines, as can be read off the $\eta \to \mu\mu$ lifetime expression in Eq.~\eqref{eq:eta_ff_nonanom}. On the other hand, the $\pi_0$ decays to $4e$ with lifetime given by Eq.~\eqref{eq:pi0_4e_nonanom}. Compared to the top panel, here we consider a smaller fine-structure constant for hidden EM. This shifts the cosmological and astrophysical constraints, resulting for example in a stronger lower bound on $\varepsilon$ from thermalization. Laboratory searches for $A'\to \mathrm{invisible}$ are essentially unaffected by this reduction in $\hat{\alpha}$, whereas the $A'\to \ell\ell$ limits become stronger, benefiting from the increased branching ratio to SM particles. In particular, the BaBar and LHCb dilepton searches currently provide the leading sensitivity in the range $8\;\mathrm{GeV} \lesssim m_{A'} \lesssim 20\;\mathrm{GeV}$, slightly outperforming EWPT and DIS.

{\it Bottom panel:} We consider a third benchmark with larger masses,
\begin{equation} \label{eq:LO_spectrum3}
m_{\pi_0, K_0, \overline{K}_0} = 300\;\mathrm{MeV}, \qquad m_{\pi_\pm, K_\pm} = 330\;\mathrm{MeV}, \qquad m_{\eta} = 339\;\mathrm{MeV},
\end{equation}
obtained by taking $B = 4\pi f_\pi = 343$~MeV, $m = 131$~MeV, and $\Delta m = 55$~MeV. The relative splittings with respect to the lightest multiplet are $\Delta_{\pi} = 0.10$ and $\Delta_{\eta} \simeq 4\Delta_\pi/3 = 0.13$. In addition, we assume $Q'$ hidden EM charges with $d_1 = 0.5$ and $\hat{\alpha} = \alpha$. In comparison with the middle panel, larger mass splittings and a shorter $\eta$ lifetime imply that there are no CMB constraints on $\eta$ decays. The isospin-breaking parameter is set to $\delta = 10^{-8}$, significantly smaller than in the top and middle panels, because here $\pi_0$ decays dominantly to $\mu\mu$ with shorter lifetime, see Eq.~\eqref{eq:pi0_ff_nonanom}. As a result, the bounds on DM decays have a different slope compared to the previous two panels, scaling like $\varepsilon \propto m_{A'}$ rather than $\varepsilon \propto m_{A'}^2$. For the chosen value of $\delta$, current Voyager~1 constraints on decaying DM rule out the entire region $m_{A'}\gtrsim 4$~GeV.

We emphasize that the values of the isospin-breaking parameter $\delta$ assumed in the three panels of Fig.~\ref{fig:summary} are simply meant to be illustrative. In all cases, changing $\delta \to \widetilde{\delta}$ rescales the exclusion and projection lines for $\pi_0$ decay by $\varepsilon\to \widetilde{\varepsilon} = \varepsilon\, (\delta / \widetilde{\delta}\,)^{1/2}$ at fixed $m_{A'}$.

\section{Summary and outlook}\label{sec:outlook}
In this paper we have studied a minimal realization of SIMP dark matter: an $SU(3)$ hidden color gauge theory with 
$N_f= 3$ light hidden quark flavors, the smallest $N_f$ that admits a Wess-Zumino-Witten action mediating $3\to 2$ self-annihilation processes. An approximate isospin $SU(2)_U$ 
global symmetry among the down and strange quarks plays a crucial role, stabilizing the lightest hidden mesons, which form a triplet of dark matter particles. The dark photon, massive and kinetically mixed with the SM hypercharge, maintains kinetic equilibrium between the hidden and visible sectors. We summarize our novel results as follows:
\begin{itemize}
    \item We performed a detailed study of the evolution and fate of the singlet meson $\eta$, which is necessarily unstable in our setup, even if the axial-vector-vector anomaly is absent in the hidden sector. We found that $\eta$ undergoes a peculiar freezeout process, driven by detailed balance between different $2\to 2$ scattering processes. Its abundance and lifetime are subject to strong constraints from CMB anisotropy measurements, but these are avoided if the meson mass splittings are larger than approximately $20\%$. 
    
    \item We studied the $3\to2$ freezeout process with mass splittings larger than the freezeout temperature, showing that the SIMP mechanism can produce the observed dark matter abundance for splittings as large as $50\%$ of the lightest meson mass. This opens up new regions of parameter space where the CMB constraints on $\eta$ decays are robustly evaded.
    
    \item We considered the possibility of decaying SIMP dark matter, as a consequence of the explicit breaking of the isospin symmetry. We analyzed indirect detection constraints in this scenario, quantifying the currently allowed amount of symmetry breaking $\delta\lesssim 10^{-5}$, and discussed future prospects.
\end{itemize}
We conclude by emphasizing a few possible directions for future work. The viable parameter space, presented in Fig.~\ref{fig:summary}, can be divided into two separate regions according to the mass of the invisibly-decaying dark photon. In the light region, $0.1\lesssim m_{A'}/\mathrm{GeV} \lesssim 5\hspace{0.2mm}$, existing or planned laboratory experiments such as Belle II and LDMX are set to test large swaths of parameter space in the future. The heavy region, $8\lesssim m_{A'}/\mathrm{GeV} \lesssim 50\hspace{0.2mm}$, appears more difficult to probe, motivating in particular further analysis of the LHC sensitivity through missing energy signatures.

As we have shown, dark matter decays can also provide powerful tests of the parameter space. Their relative importance can be put on firmer ground in the context of concrete ultraviolet completions, where a preferred size of $\delta$ may be predicted. As a step in this direction, we have outlined an embedding of our setup in the neutral naturalness framework, where a hidden $SU(3)$ gauge theory with GeV-scale confinement is rather generic. We have only sketched the main guidelines, pointing out that ingredients beyond the minimal models are required, whereas the construction of a full completion and its detailed study are left for future work. 

Finally, our dark matter $SU(2)_U$ triplet has a rather large self-scattering cross section, mediated by the kinetic term of the nonlinear sigma model. As is well-known, this cross section can be reduced in scenarios where the dark matter is composed of a single real meson species, whose self-interactions are then mediated by smaller explicit symmetry breaking effects. We have found, however, that this is not straightforward to realize in practice: a sufficient mass splitting of the triplet components requires values of $\delta$ that are in stark conflict with bounds on dark matter decay. We believe this aspect deserves further attention.   

\acknowledgments We thank Hsin-Chia~Cheng, Marco Cirelli, Admir~Greljo, Ulrich~Haisch, Simon~Knapen, Ranjan~Laha, Robert~Szafron, Yuhsin~Tsai for helpful conversations, and Yonit Hochberg and Eric Kuflik for a clarification regarding Ref.~\cite{Hochberg:2018vdo}. 

\appendix

\section{Boltzmann equations}\label{sec:BEs}
The $5$-meson interaction in the last line of Eq.~\eqref{eq:ChPT} can be written in the form~\cite{Hochberg:2014kqa}
\begin{equation}
\frac{N_c}{240\pi^2 f_\pi^5}\, \epsilon^{\mu\nu\rho\sigma} \sum_{a \,<\, b\, <\, c\, <\, d\, <\, e} T_{abcde} \pi^a \partial_\mu \pi^b \partial_\nu \pi^c \partial_\rho \pi^d \partial_\sigma \pi^e \,,
\end{equation}
where $T_{abcde} = 60\,\mathrm{Tr}(\lambda^a \lambda^b \lambda^c \lambda^d \lambda^e)$, and explicit calculation gives
\begin{align}
&\qquad T_{12345} = T_{12367} = - 60\,, \qquad T_{45678} = 40 \sqrt{3}\,, \nonumber \\ T_{12458} =& - T_{12678} = - T_{13468} = - T_{13578} = T_{23478} = - T_{23568} = - 20 \sqrt{3}\,,
\end{align}
while all other entries vanish. Defining $T_{\{abcde\}}$ as the ordered form, i.e. $T_{\{31542\}} = T_{12345}$ etc., we find $t^{2} \equiv \frac{1}{5!^2} \sum T_{\{abcde\}}^2 = 160$, in agreement with the general expression $t^2 = 4 N_f (N_f^2 - 1) (N_f^2 -4)/3$ given in Ref.~\cite{Hochberg:2014kqa}. The general form of the Boltzmann equations that control the evolution of the meson yields is, using $x = m_{\pi_0}/T$ as the time variable,
\begin{align}
\frac{dY_a}{dx} = -\,& \frac{\lambda}{x^5 3!2!} \sum_{b,\,c,\,d,\,e}   \frac{\langle\sigma v^2 \rangle_0 T^2_{\{abcde\}}}{(5!)^2} \Big(Y_a Y_b Y_c - Y_a^{\rm eq} Y_b^{\rm eq} Y_c^{\rm eq} \frac{Y_d Y_e}{Y_d^{\rm eq} Y_e^{\rm eq}} \Big)\nonumber \\
-\,& \frac{\kappa}{x^2 (2!)^2} \sum_{b,\,c,\,d} \,\langle \sigma v \rangle_{ab\to cd}\, \big(c_{abcd} + \tfrac{1}{3} R_{abcd}\big)^2 \Big(Y_a Y_b - Y_a^{\rm eq} Y_b^{\rm eq} \frac{Y_c Y_d}{Y_c^{\rm eq} Y_d^{\rm eq}} \Big)\,,
\end{align}
where we have defined the quantities
\begin{align}
\lambda \equiv \frac{4\sqrt{10}\, \pi^3}{675} \frac{M_{\rm Pl}m_{\pi_0}^4 g_{\ast s}^2}{\sqrt{g_\ast}}\,, \qquad \kappa \equiv \frac{2\sqrt{10}\, \pi}{15} \frac{g_{\ast s} M_{\rm Pl} m_{\pi_0}}{\sqrt{g_\ast}}\,,
\end{align}
with $M_{\rm Pl}$ being the reduced Planck mass, while $\langle \sigma v^2 \rangle_0$ was defined in Eq.~\eqref{eq:sigmav2_def}. The thermally-averaged cross section for $2\to 2$ scattering is
\begin{equation} \label{eq:2to2_sigmav}
\langle \sigma v \rangle_{ab\, \to\, cd} \simeq \frac{m_{\pi_0}^2}{128 \pi f_\pi^4} \,\beta_{abcd}\,,\qquad \beta_{abcd} \equiv \sqrt{1 - 2\, \frac{ m_c^2 + m_d^2}{(m_a + m_b)^2} + \frac{ (m_c^2 - m_d^2)^2 } {(m_a + m_b)^4} }\;,
\end{equation}
while $c_{abcd}$ and $R_{abcd}$ were defined after Eq.~\eqref{eq:self_int}.

We now present the explicit Boltzmann equations for the three multiplets, written in terms of the yields per degree of freedom $Y_i$ ($i = \eta, \pi_+, \pi_0$). We have
\begin{align}
\frac{dY_{\eta}}{dx} =\,& - \frac{\lambda \langle \sigma v^2 \rangle_0}{x^5 3! 2!\hspace{0.15mm} 3 } \bigg[ 6 \Big( Y_\eta Y_+^2 - Y_\eta^{\rm eq} (Y_{+}^{\rm eq})^2 \frac{Y_0^2}{(Y_0^{\rm eq})^2} \Big) + 24 \Big( Y_\eta Y_+ Y_0 - Y_\eta^{\rm eq} Y_+ Y_0 \Big) \nonumber \\
\,&\qquad\qquad\qquad\qquad+ 6 \Big( Y_\eta Y_0^2 - Y_\eta^{\rm eq} (Y_{0}^{\rm eq})^2 \frac{Y_+^2}{(Y_+^{\rm eq})^2} \Big) 
+\,24 \Big( Y_\eta Y^2_+ - Y_\eta^{\rm eq} Y_+^2 \Big) \bigg] \nonumber \\
\,- \frac{\kappa}{x^2 (2!)^2 9}& \bigg[ 784 \langle \sigma v \rangle_{HH\to MM}  \Big( Y_\eta^2 - (Y_\eta^{\rm eq})^2 \frac{Y_+^2}{(Y_+^{\rm eq})^2} \Big) +  12 \langle \sigma v \rangle_{HH\to LL} \Big( Y_\eta^2 - (Y_\eta^{\rm eq})^2 \frac{Y_0^2}{(Y_0^{\rm eq})^2} \Big)   \nonumber \\
\,+ 288\,& \langle \sigma v \rangle_{HM\to ML} \Big( Y_\eta Y_+ - Y_\eta^{\rm eq} \frac{Y_+ Y_0}{Y_0^{\rm eq}} \Big) -  576\, \langle \sigma v \rangle_{MM\to HL} \Big( Y_+^2 - (Y_+^{\rm eq})^2 \frac{Y_\eta Y_0}{Y_\eta^{\rm eq}Y_0^{\rm eq}} \Big) \bigg]\,, \label{eq:Yeta}
\end{align}
\begin{align}
\frac{dY_{+}}{dx} =\,& - \frac{\lambda \langle \sigma v^2 \rangle_0 }{ x^5 3! 2! \hspace{0.15mm} 3} \bigg[ 9 \Big( Y_+^2 Y_0 - (Y_+^{\rm eq})^2 \frac{Y_0^2}{Y_0^{\rm eq}} \Big) + 9 \Big( Y_+ Y_0^2 - Y_0^{\rm eq}Y_+ Y_0 \Big) \nonumber \\ 
+&\, 3 \Big( Y_+^2 Y_\eta - (Y_+^{\rm eq})^2 Y_\eta^{\rm eq} \frac{Y_0^2}{(Y_0^{\rm eq})^2} \Big) 
+ 6 \Big( Y_+^2 Y_0 - (Y_+^{\rm eq})^2 \frac{Y_0 Y_\eta}{Y_\eta^{\rm eq}} \Big) + 6 \Big( Y_+ Y_\eta Y_0 - Y_\eta^{\rm eq} Y_+ Y_0 \Big)\nonumber \\ 
+&\, 3 \Big( Y_+ Y_0^2 - (Y_0^{\rm eq})^2 \frac{Y_+ Y_\eta}{Y_\eta^{\rm eq}} \Big)
  + 12 \Big( Y_+^3  - (Y_+^{\rm eq})^2 \frac{Y_+ Y_\eta}{Y_\eta^{\rm eq}} \Big) 
+12 \Big( Y_+^2 Y_\eta  - Y_\eta^{\rm eq} Y_+^2 \Big) \bigg] \nonumber \\
\,-& \frac{\kappa}{ x^2 (2!)^2 9} \bigg[ - 196 \langle \sigma v \rangle_{HH\to MM}  \Big( Y_\eta^2 - (Y_\eta^{\rm eq})^2 \frac{Y_+^2}{(Y_+^{\rm eq})^2} \Big) \nonumber \\
\,+288\,& \langle \sigma v \rangle_{MM\to HL}  \Big( Y_+^2 - (Y_+^{\rm eq})^2 \frac{Y_\eta Y_0 }{Y_\eta^{\rm eq} Y_0^{\rm eq} } \Big) 
+108\, \langle \sigma v \rangle_{MM\to LL} \Big( Y_+^2 - (Y_+^{\rm eq})^2 \frac{Y_0^2}{(Y_0^{\rm eq})^2} \Big) \bigg]\,,  \label{eq:Ypiplus}
\end{align}
\begin{align}
\frac{dY_{0}}{dx} =\,& - \frac{\lambda \langle \sigma v^2 \rangle_0 }{ x^5 3! 2!  \hspace{0.15mm} 3} \bigg[ 6 \Big( Y_0^3 - (Y_0^{\rm eq})^3 \frac{Y_+^2}{(Y_+^{\rm eq})^2} \Big) + 6 \Big( Y_0 Y_+^2 - (Y_+^{\rm eq})^2 \frac{Y_0^2}{Y_0^{\rm eq}} \Big) \nonumber \\ 
\,&+ 24 \Big( Y_0^2 Y_+ - Y_0^{\rm eq} Y_+ Y_0 \Big)    
+ 8  \Big( Y_0 Y_+ Y_\eta - Y_\eta^{\rm eq} Y_+ Y_0 \Big) +4 \Big( Y_0 Y_+^2 - (Y_+^{\rm eq})^2 \frac{Y_0 Y_\eta}{Y_\eta^{\rm eq}} \Big)\nonumber \\ 
\,&+ 8 \Big( Y_0^2 Y_+  - (Y_0^{\rm eq})^2 \frac{ Y_+ Y_\eta}{Y_\eta^{\rm eq}} \Big)  
 + 4 \Big( Y_0^2 Y_\eta  - (Y_0^{\rm eq})^2 Y_\eta^{\rm eq} \frac{ Y_+^2}{(Y_+^{\rm eq})^2} \Big)  \bigg] \nonumber \\
\, - \frac{\kappa}{ x^2 (2!)^2 9}& \bigg[ - 4 \langle \sigma v \rangle_{HH\to LL} \Big( Y_\eta^2 - (Y_\eta^{\rm eq})^2 \frac{Y_0^2}{(Y_0^{\rm eq})^2} \Big)
 - 96\, \langle \sigma v \rangle_{HM\to ML} \Big( Y_\eta Y_+ - Y_\eta^{\rm eq} \frac{Y_+ Y_0}{Y_0^{\rm eq}} \Big) \nonumber \\
\,- 192\,& \langle \sigma v \rangle_{MM\to HL} \Big( Y_+^2 - (Y_+^{\rm eq})^2 \frac{Y_\eta Y_0}{Y_\eta^{\rm eq}Y_0^{\rm eq}} \Big) 
- 144\, \langle \sigma v \rangle_{MM\to LL} \Big( Y_+^2 - (Y_+^{\rm eq})^2 \frac{Y_0^2}{(Y_0^{\rm eq})^2} \Big) \bigg]\,, \label{eq:Ypi0}
\end{align}
where we have defined $Y_{+} \equiv Y_{\pi_+}$ and $Y_0 \equiv Y_{\pi_0}$. We have assumed that all $3\to 2$ annihilations are kinematically allowed at $T = 0$. We have also taken $2 m_{\pi_+} > m_\eta + m_{\pi_0}$, as verified for our leading-order spectrum. The equilibrium yields are defined as $Y_i^{\rm eq}(x) = Y^{\rm eq}(x\hspace{0.2mm} m_i/m_{\pi_0})$, where we employ the non-relativistic approximation
\begin{equation}
Y^{\rm eq}(z) = \frac{45}{4\pi^4}\frac{g}{g_{\ast s}} z^2 K_2(z)\,,\qquad g = 1\,.    
\end{equation}
In addition, $g_\ast$ and $g_{\ast s}$ are also $x$-dependent.

The above assumes that the hidden and SM sectors are in kinetic equilibrium at temperature $T$. If kinetic decoupling occurs at $x_{\rm dec}$ (see e.g. the dotted curves in the left panel of Fig.~\ref{fig:Boltzmann} and in Fig.~\ref{fig:Boltzmann_2}, which correspond to $x_{\rm dec} = 25 \gtrsim x_{\rm fo}^{3\to 2}$), for $x>x_{\rm dec}$ we solve Eqs.~(\ref{eq:Yeta}$\,$--$\,$\ref{eq:Ypi0}) including only $2\to 2$ processes, and with equilibrium yields now given by $Y_i^{\rm eq}(x) = Y_\xi^{\rm eq}(x\hspace{0.2mm} m_i/m_{\pi_0})$, where
\begin{equation}
Y_\xi^{\rm eq}(z) = \frac{45}{4\pi^4}\frac{g}{g_{\ast s}} \xi z^2 K_2(z/\xi)\,,\qquad g = 1\,.   
\end{equation}
Here $\xi \equiv T_D/T$ is the ratio of temperatures, given by $\xi = x_{\rm dec}/x$ at $x > x_{\rm dec}$.

It can be checked that the evolution of the total meson yield $Y_\pi \equiv Y_\eta + 4 Y_+ + 3 Y_0$ is unaffected by $2\to 2$ scatterings. In addition, in the limit of degenerate masses, $Y_\pi$ satisfies
\begin{equation}
\frac{dY_\pi}{dx} = - \frac{\lambda}{x^5} \langle \sigma v^2 \rangle_{3\to 2} (Y_\pi^3 - Y_\pi^{\rm eq} Y_\pi^2)\,,\qquad  \langle \sigma v^2 \rangle_{3\to 2} =  \frac{5\sqrt{5}\,m_\pi^5 N_c^2}{2\pi^5 2^{10} f_\pi^{10} x^2} \frac{t^2}{N_\pi^3}
\end{equation}
as in Ref.~\cite{Hochberg:2014kqa}, where $t^2$ was given above and $N_\pi = N_f^2 - 1$. Note that our normalization of the pion decay constant differs from that in Ref.~\cite{Hochberg:2014kqa} by a factor of $2$.

\bibliographystyle{JHEP}
\bibliography{pionlit}

\end{document}